\documentclass{article}
\usepackage{lscape}
\usepackage{graphicx}
\usepackage{subcaption}
\usepackage{float} 
\usepackage{caption}
\usepackage{PRIMEarxiv}
 \usepackage{amsmath}
\usepackage[utf8]{inputenc} 
\usepackage[T1]{fontenc}    
\usepackage{hyperref}       
\usepackage{url}            
\usepackage{booktabs}       
\usepackage{amsfonts}       
\usepackage{nicefrac}       
\usepackage{microtype}      
\usepackage{xspace}
\newcommand{\pkg}[1]{\texttt{#1}\xspace}
\usepackage{lipsum}
\usepackage{bm}             
\usepackage{fancyhdr}       
\usepackage{graphicx}       
\graphicspath{{media/}}     
\usepackage{natbib}
\usepackage{comment, xcolor}
\usepackage{subcaption}

\title{A critical comparison of handling zeros in high-dimensional  compositional count data\thanks{\textit{\underline{Citation}}: 
\textbf{Authors. Title. Pages.... DOI:000000/11111.}} 
}

\author{
  Wenqi Tang \\
  Department of Mathematics and Statistics\\ University of Jyv\"askyl\"a\\ 
  Finland\\
  \texttt{wenqi.t.tang@jyu.fi} \\
   \And
   Kamila Fa\v cevicov\'a \\  
Department of Mathematical Analysis and \\ Applications of Mathematics\\ Palack\'y University Olomouc\\
Czech Republic\\
  \texttt{kamila.facevicova@upol.cz} \\
   \AND
  Klaus Nordhausen \\
  Department of Mathematics and Statistics\\ University of Helsinki\\ 
  Finland\\
  \texttt{klaus.nordhausen@helsinki.fi} \\
   \And
  Sara Taskinen \\
  Department of Mathematics and Statistics\\ University of Jyv\"askyl\"a\\ 
  Finland\\
  \texttt{sara.l.taskinen@jyu.fi} \\
}

\begin{document}
\maketitle

\begin{abstract}

The growing use of high-throughput sequencing (HTS) has enabled the large-scale production of compositional count data, driving progress in microbiome research. However, such count data are often high-dimensional, over-dispersed, and heavily zero-inflated, and they conflict with the continuity assumptions underlying log-ratio–based compositional data analysis (CoDA), creating substantial methodological challenges.
This review provides a overview of zero-handling strategies in compositional data, covering zero-tolerant transformations, imputation approaches for rounded zeros, and statistical models for essential zeros. We specifically highlight the problems that arise when applying the log-ratio framework to sequencing-derived compositional count data, where violations of continuity can induce numerical instabilities and biased statistical inferences. Motivated by these issues, we systematically examine how existing imputation strategies behave when adapted to discrete, zero-inflated count data, including an evaluation of how the discrete, lattice-valued nature of the data affects imputation performance.

Overall, this review consolidates scattered methodological developments, clarifies appropriate use cases, and identifies open challenges that motivate future zero-handling frameworks capable of jointly accommodating compositional constraints, zero inflation, and the lattice nature of count data, while also providing a detailed discussion of the comparison results.
\end{abstract}

\keywords{Compositional data analysis \and Zeros  \and High-throughput sequencing (HTS) \and Microbiome count data \and Imputation}

\section{Introduction}
\label{sec:intro} 

Recent advances in modern sampling and classification techniques, such as high-throughput sequencing (HTS) \citep{Fernandes2013, Conesa2016, Pollock2018}, which allow rapid and large-scale generation of DNA or RNA sequence data \citep{GloorMacklaimPawlowskyGlahnEgozcue2017}, have greatly advanced microbiome research. The data obtained through these techniques typically consist of thousands to millions of sequencing reads per sample, representing multiple taxa within a microbial community. In large-scale studies, the total number of reads can reach hundreds of millions across all samples \citep{DiBella2013}. These read count tables, which quantify the number of sequences assigned to microbial taxa or genes, form the basis for downstream analyses linking microbial composition to host phenotypes, environmental factors, and clinical outcomes.

The sequencing approaches enable partitioning of HTS data into biologically meaningful units. In practical analyses, microbiome data are often represented as operational taxonomic units (OTUs) or amplicon sequence variants (ASVs), which consist of relative counts and exhibit unique statistical characteristics that pose significant analytical challenges. First, microbiome count data are typically high-dimensional, as the number of taxa or features can easily exceed thousands even for a modest number of samples. Second, the data are highly sparse, with many zero or near-zero values resulting from either limited sequencing depth or the true absence of taxa. Finally, sequencing data are inherently compositional because sequencing instruments have a fixed capacity: the total number of reads obtained from a sample is distributed among all observed sequences, meaning that only relative information is preserved \citep{GloorMacklaimPawlowskyGlahnEgozcue2017}.

As a result, microbiome count data are inherently compositional, high-dimensional, and sparse, further complicated by technical variation and sequencing depth differences. These features make standard statistical methods inappropriate when applied directly to raw sequencing data without suitable normalization or transformation. To address the relative nature of compositional data, \citet{Aitchison1986} introduced a comprehensive log-ratio framework that focuses on ratios between components rather than their absolute values (see Sec.~\ref{sec:composition}).  However, when applied to microbiome count data, several additional challenges arise.
The most fundamental issue is the prevalence of zeros: extreme sparsity—common in
microbiome sequencing data and, in some  cases, approaching 90\% zeros
\citep{GloorMacklaimPawlowskyGlahnEgozcue2017}—
precludes the direct use of log-ratio transformations (see Sec.~\ref{sec: trans}),
which are otherwise essential for mapping simplex-constrained data into a Euclidean
space suitable for conventional statistical analyses. 
Beyond zero inflation, the discrete and lattice-valued nature of count data violates
the continuity assumptions underlying the log-ratio framework, introducing quantization effects that can strongly affect small values and distort analytical results \citep{LovellChuaMcGrath2020, BaconShone2008}.
Although this review focuses primarily on addressing zeros, these discreteness-related
considerations further underscore the need for caution when applying CoDA methods to
count-based microbiome data.

The layout of this article is as follows. Section~\ref{sec:composition} provides a brief introduction to the foundations of compositional data analysis (CoDA) and the role of log-ratio transformations. The main review is in Sec ~\ref{sec:rev} and figure~\ref{fig:review_outline} presents an overview of the overall structure of the review. Before discussing strategies for handling zeros, we first introduce the different types of zeros, their underlying causes, and the principles that guide their appropriate treatment, emphasising the necessity of mechanism-based zero handling (Section~\ref{sec:zero_type}). Section~\ref{sec:rounding} focuses on strategies for handling rounded zeros—those arising from detection limits or insufficient sampling—and reviews a range of imputation methods developed for constructing strictly positive compositions. Section~\ref{sec:essential} then turns to essential zeros, providing an overview of modelling frameworks—such as mixture-based logistic-normal models, latent-Gaussian approaches, and Dirichlet-type formulations—that preserve the structural meaning of true absences rather than replacing them. Section~\ref{sec:trans0} introduces zero-tolerant transformations, which depart from conventional log-ratio transformations and enable zeros to be incorporated directly into statistical analysis, including power-based and directional transformations. Given that our motivating datasets are high-dimensional, sparse, and derived from sequencing counts, Section~\ref{sec:count} is dedicated to the specific challenges posed by discrete count compositional data. We discuss how the lattice-valued nature of sequencing counts violates the continuity assumptions underlying classical CoDA and review current strategies tailored to discrete compositions, including imputation approaches and probabilistic models such as Dirichlet–multinomial and zero-inflated formulations. 

Building on the theoretical considerations in Section~\ref{sec:count}, Section~\ref{sec:comp} presents a comparative evaluation of existing imputation methods using real high-dimensional count-based datasets. We first acknowledge the important benchmark provided by \cite{LubbeFilzmoserTempl2021}, which compared classical and machine-learning zero-replacement techniques. Their study utilized real microbiome data but treated them as continuous compositions—interpreting sequencing-derived zeros as detection-limit issues and replacing them with a fixed value (set to 1), without explicitly accounting for their discrete, count-based origin. In stark contrast, our comparison explicitly addresses the specific constraints of count-based data, whose wide dynamic ranges, inherent discreteness, and heterogeneous sequencing depths fundamentally affect imputation behaviour. Since analysts often require a method compatible with a broad range of standard multivariate techniques (e.g., PCA, clustering), our primary goal is to evaluate how imputation methods restore completeness and covariance structure and support general downstream multivariate analyses in a model-agnostic manner. Consequently, we exclude methods typically embedded within specific modeling structures, as they require parameter optimization tied to particular statistical models.
To achieve an objective assessment, we employ a fully observed sequencing-derived count dataset, the Rabbit dataset \citep{greenacre2021compositional}, which exhibits the typical characteristics of high-dimensional microbial compositions (discreteness, zero inflation, overdispersion, and strong imbalance). By introducing zeros under controlled conditions and comparing imputed values with known true counts, we obtain a direct and objective assessment of imputation accuracy. The experimental design consists of two parts that explicitly quantify the influence of both zero proportion and dimensionality, thereby providing a comprehensive evaluation across multiple sparsity and dimensionality regimes. An identical design is also applied to simulated data with matching compositional and discrete characteristics, as presented in Appendix~\ref{appendix.A}, further corroborating the robustness of our conclusions.

Our review covers the key conceptual and methodological contributions of this work as follow. First, this review provides the first unified, mechanism-based framework for zero handling in compositional data, integrating replacement strategies, model-based approaches, and zero-tolerant transformations into a coherent structure. Second, it clarifies the fundamental mismatch between continuous CoDA theory and discrete count data, integrating theoretical insights and empirical evidence to show why HTS-derived compositions—with their inherent lattice structure and scale-dependent properties—require considerations beyond classical log-ratio methods. Third, the review uses a systematic comparative evaluation using real and simulated high-dimensional count datasets, offering evidence-based guidance for method selection under varying sparsity, dimensionality, and scale. The results of this comparative study simultaneously reveal where continuous-data imputation methods underperform due to numerical instabilities caused by discreteness, and where discrete-aware adjustments become beneficial. Collectively, these contributions establish a methodological foundation that exposes the gap between existing continuous-data approaches and the specific requirements of count-based applications, laying the groundwork for future developments that jointly respect zero inflation, compositional constraints, and the lattice-valued nature of sequencing-derived count data.

\section{Compositional Data Analysis}
\label{sec:composition}
Compositional data, or compositions, are vectors of non-negative values that describe the contribution of $D$ parts constrained by a constant sum, typically 1 (for proportions) or 100 (for percentages). 
For a $D$-part composition $\mathbf{x} = (x_1, \dots, x_D)'$ lies in a $(D-1)$-dimensional simplex space $\mathcal{S}^D$:  
\[
\mathcal{S}^D = \left\{ \mathbf{x} = (x_1, \dots, x_D)' \,\bigg|\, x_i > 0, \, \sum_{i=1}^D x_i = C \right\}.
\]  
The right panel of Fig~\ref{fig:lattice-simplex} shows that when $D=3$, a composition resides in the 
continuous and infinitely dense simplex $\mathcal{S}^3$, which is geometrically 
represented as a triangular plane containing arbitrarily many intermediate points

Due to the constant-sum constraint, compositional components are inherently dependent—an increase in one part necessarily decreases others—making conventional statistical methods inappropriate. Such methods assume Euclidean geometry and overlook the relative nature and interdependence among parts, often leading to spurious correlations.  To address these issues, \citet{Aitchison1986} established the foundation of Aitchison geometry, which defines compositional data analysis (CoDA) through three fundamental principles: scale invariance, permutation invariance, and subcompositional coherence. Scale invariance ensures that multiplying all parts of a composition by a positive constant does not change the information it conveys. Permutation invariance guarantees that results are unaffected by the ordering of components. Subcompositional coherence requires that analyses based on a subset of components yield conclusions consistent with those from the full composition.  These principles form the theoretical basis for the log-ratio framework, the cornerstone of modern CoDA, providing a robust and interpretable foundation for analyzing compositional data.   

\subsection{Log-ratio framework}

In practice, the requirement in the traditional definition that compositional parts must add up to a fixed constant sum $C$ introduces several challenges. First, the presence of zeros is problematic: missing or unmeasured components may violate the sum constraint, and rounding errors can cause deviations from the prescribed total. Second, although many real-world datasets exhibit a compositional character, their total sum often depends on sample-specific technical factors rather than meaningful biological quantities. For example, in microbiome data, the total read count of each sample is determined largely by sequencing depth, which reflects technical variation. As emphasized by \citet{GloorMacklaimPawlowskyGlahnEgozcue2017}, sequencing count tables are compositional, but their total read counts have no biological interpretation and therefore cannot be used to compare samples on an absolute scale.

Under the constant-sum constraint, the values of individual parts are inherently interdependent—an increase in one component necessarily affects the others. Therefore, CoDA focuses on analyzing ratios between components rather than their absolute values or total counts, as the ratios more accurately capture the relative structure of the data. \citet{FilzmoserHronTempl2018} emphasized that compositional data describe relative information among components, and that proportions just represent only one possible expression of this information, not its defining characteristic.
Under scale invariance, multiplying a composition $\mathbf{x}$ by any positive constant $c>0$ alters the total sum but leaves all ratios $x_i/x_j$ unchanged; hence $\mathbf{x}$ and $c\mathbf{x}$ convey the same relative information. It is worth noting that scale invariance fundamentally relies on the continuous nature of the simplex. However, when the data are discrete lattice-valued counts, this property breaks down: scaling produces non-integer values that, if we wish to preserve discreteness, must be mapped back onto the lattice through rounding, thereby introducing quantization error that distorts the original ratios. We return to this issue in detail in Sec.~\ref{sec:count}.

Building on this idea, \citet{Aitchison1986} developed the CoDA framework based on log-ratio methodology, which represents compositional data in terms of their relative structure rather than absolute magnitudes. For example, this framework defines a meaningful distance between two compositions $\mathbf{x}$ and $\mathbf{y} \in \mathcal{S}^D$, known as the Aitchison distance:  
\begin{equation}
d_A(\mathbf{x}, \mathbf{y}) = 
\sqrt{\frac{1}{2D} \sum_{i=1}^{D} \sum_{j=1}^{D} 
\left( \ln \frac{x_i}{x_j} - \ln \frac{y_i}{y_j} \right)^2 }.
\label{eq:ait_distance}
\end{equation}

\subsection{Transformations in CoDA}
\label{sec: trans}
As discussed above, compositional data, constrained to the Aitchison simplex space $\mathcal{S}^D$, exhibit several critical issues. 
First, the interdependence among parts induces spurious correlations. 
Second, for a composition with $D$ parts, only $D-1$ parts are free to vary, as the last component is determined by the constant-sum constraint, effectively reducing the degrees of freedom. 
Finally, most classical statistical methods, including correlation and regression analysis, assume unconstrained data; applying them directly to compositional data can yield misleading or entirely incorrect results.

To address these challenges, CoDA applies log-ratio transformations to map data from the constrained simplex space $\mathcal{S}^D$ to an unconstrained Euclidean space. By focusing on ratios between components rather than their absolute values, these transformations eliminate distortions caused by the constant-sum constraint and enable the use of standard statistical techniques. Common implementations include the additive logratio (ALR), centered logratio (CLR), and isometric logratio (ILR) transformations. ALR facilitates pairwise comparisons by taking the logarithm of each component relative to a selected reference, CLR preserves symmetry by using the geometric mean of all components as the denominator, and ILR constructs orthogonal balances to maintain geometric properties for multivariate analyses. Each transformation serves a specific analytical purpose, with the choice depending on interpretability and study objectives \citep{FilzmoserHronTempl2018}. However, because all rely on logarithmic operations, none can directly accommodate zero values—a central issue underlying the compositional zero problem.

\subsubsection{Additive log-ratio transformation (ALR)}
The ALR transformation maps a composition 
$\mathbf{x} = (x_{1}, \dots, x_{D})'$ 
from the simplex space $\mathcal{S}^D$ to $\mathbb{R}^{D-1}$ by selecting a reference component $x_j$, as follows:
\begin{equation}
\mathbf{z} = \text{ALR}_j(\mathbf{x}) = \left( 
\ln \frac{x_1}{x_j}, \dots, \ln \frac{x_{j-1}}{x_j}, \ln \frac{x_{j+1}}{x_j}, \dots, \ln \frac{x_D}{x_j} 
\right)'.
\label{ALR}
\end{equation}

Here, $x_j$ is a pre-selected reference component. ALR-transformed data $\mathbf{z}$ directly reflect whether the remaining components increase or decrease relative to the reference component $x_j$. For example, in biological applications, if a biologically stable and ubiquitous species (such as certain core microbiota in microbiome studies) is chosen as the reference, the results can have clear biological significance. \citet{Zhang2024} suggested that an ideal reference component should be interpretable, highly abundant, stable, prevalent across all samples (with few zeros), show no significant differences between groups (i.e., not itself a target of study), and exhibit minimal variance in relative abundance after log transformation. When a clear and stable reference is available, the ALR transformation is computationally simple and highly interpretable.

 It should be noted that the ALR transformation is not isometric, which means that  $d_A(\mathbf{x}, \mathbf{y}) \neq d_E(\text{ALR}(\mathbf{x}), \text{ALR}(\mathbf{y}))$,
where $d_A$ denotes the Aitchison distance defined in Eq.~\ref{eq:ait_distance}  and $d_E$ the Euclidean distance \citep{Aitchison1986}. Consequently, to ensure consistent interpretation, statistical methods applied to ALR-transformed data should be invariant under any permutation of components. Due to the requirement of choosing a reference part, ALR coordinates provide relative information only with respect to that component, and outliers in the reference can propagate distortions across all ratios, potentially biasing results. 

Building on the ALR transformation, \citet{Aitchison1986} introduced the additive logistic normal (ALN) distribution, which offers a probabilistic framework for modeling compositional data. Specifically, a composition $\mathbf{x} \in \mathcal{S}^D$ follows an ALN distribution if its ALR-transformed vector follows a multivariate normal distribution $\mathbf{z} \sim \mathcal{N}_{D-1}(\boldsymbol{\mu}, \boldsymbol{\Sigma})$ or equivalently, $\mathbf{x} \sim \text{ALN}_D(\boldsymbol{\mu}, \boldsymbol{\Sigma})$, where $\boldsymbol{\mu}$ and $\boldsymbol{\Sigma}$ denote the mean vector and covariance matrix in the ALR-transformed space.

\subsubsection{Central log-ratio transformation (CLR)}
The CLR transformation maps a composition $\mathbf{x}$ from the simplex space $\mathcal{S}^D$ into a $(D - 1)$-dimensional hyperplane in $\mathbb{R}^D$ as follows:
\[
\mathbf{z} = \text{CLR}(\mathbf{x}) = \left( z_1, \dots, z_D \right)^{\prime}
= \left( \ln \frac{x_1}{\sqrt[D]{\prod_{i=1}^D x_i}}, \dots, 
\ln \frac{x_D}{\sqrt[D]{\prod_{i=1}^D x_i}} \right)^{\prime}.
\]

Here, the CLR transformation uses the geometric mean as the denominator. Compared with the ALR transformation, which requires a preselection of a reference component, CLR-transformed data are symmetric and unique, maximizing the preservation of relative information among components. Additionally, the CLR transformation is isometric, meaning that the geometric relationships (distances) between compositions in the original simplex are preserved in the transformed space.  

After the CLR transformation, the data are projected onto a hyperplane where the sum of all components is zero. This allows standard multivariate statistical methods based on Euclidean distance to be applied. However, it is important to note that the CLR-transformed variables sum to zero, which implies linear dependence among the $D$ new variables. This leads to a singular covariance matrix, complicating the application of many multivariate statistical methods (such as those requiring matrix inversion) and potentially resulting in unreliable outcomes \citep{FilzmoserHronTempl2018}.  

\subsubsection{Isometric log-ratio transformation (ILR)}
The ILR transformation maps a composition 
$\mathbf{x} = (x_{1}, \dots, x_{D})'$ 
from the simplex space $\mathcal{S}^D$ to $\mathbb{R}^{D-1}$. 
It is derived from the CLR-transformed data by applying a given orthogonal matrix $H$, that is,  
\[
\mathbf{z} = \text{ILR}(\mathbf{x}) = H' \, \text{CLR}(\mathbf{x}),
\]
where $H$ is a $D \times (D-1)$ orthonormal contrast matrix, typically obtained as a submatrix of the Helmert matrix \citep{boogaart2013analyzing}.  
The resulting ILR coordinates strictly preserve distances and consist of $D-1$ independent variables, thereby resolving the collinearity issue inherent in CLR-transformed data.

Each ILR coordinate usually involves multiple compositional parts, making interpretation less straightforward \citep{boogaart2013analyzing}.  
To facilitate interpretability, a generalized system of pivot coordinates has been proposed \citep{FilzmoserHronTempl2018}, in which $D$ alternative ILR transformations are constructed such that, in each case, the first coordinate captures the relative information of a specific part by permuting it to the first position \citep{FiserovaHron2011}.

To construct pivot coordinates, the $l$-th part of composition $\mathbf{x}$ is moved to the first position,
\[
\mathbf{x}^{(l)} = (x_l, x_1, \dots, x_{l-1}, x_{l+1}, \dots, x_D)^{\prime}.
\]

Applying the ILR transformation with a Helmert-based orthonormal basis yields
\[
z_i^{(l)} = \sqrt{\frac{D-i}{D-i+1}} 
\ln \frac{x_i^{(l)}}{\left(\prod_{j=i+1}^{D} x_j^{(l)}\right)^{1/(D-i)}},
\quad i = 1,\dots,D-1.
\]

The key property 
is that the first coordinate $z_1^{(l)}$ contains all relative information
about part $x_l$ with respect to the remaining parts, while the remaining
coordinates $z_2^{(l)}, \dots, z_{D-1}^{(l)}$ describe the log-ratios among the other components
\citep{FiserovaHron2011}. The ILR coordinates offer a robust and interpretable framework that preserves the geometric structure of the data, making them particularly useful for statistical modeling and inference in compositional data analysis \citep{FilzmoserHronTempl2018}.

However, the strong reliance on the logarithmic function across all standard log-ratio transformations (ALR, CLR, and ILR) necessitates that all components $\mathbf{x}$ be strictly positive. Consequently, these transformations cannot directly accommodate any zero values---a central issue underlying the compositional zero problem. In the subsequent Sec.~\ref{sec:rev}, we briefly introduce the main existing strategies for handling zeros in compositional data. These approaches can be grouped into three major lines: (i) imputing the missing parts to obtain a full data matrix (Sec.~\ref{sec:rounding}); (ii) constructing statistical models that explicitly account for the meaning and mechanism of zeros (Sec.~\ref{sec:essential}); and (iii) developing zero-tolerant transformations (Sec.~\ref{sec:trans0}), such as the $\alpha$-transformation---which aim to achieve a practical balance between the strict geometric principles of Log-Ratio Analysis (LRA) and the simplicity of Raw Data Analysis (RDA), thereby enhancing feasibility for zero-inflated and high-dimensional sparse compositions.

\section{Review of methods dealing with zeros in CoDA}

\label{sec:rev}
\begin{figure}[ht]
    \centering
    \includegraphics[width=1\linewidth]{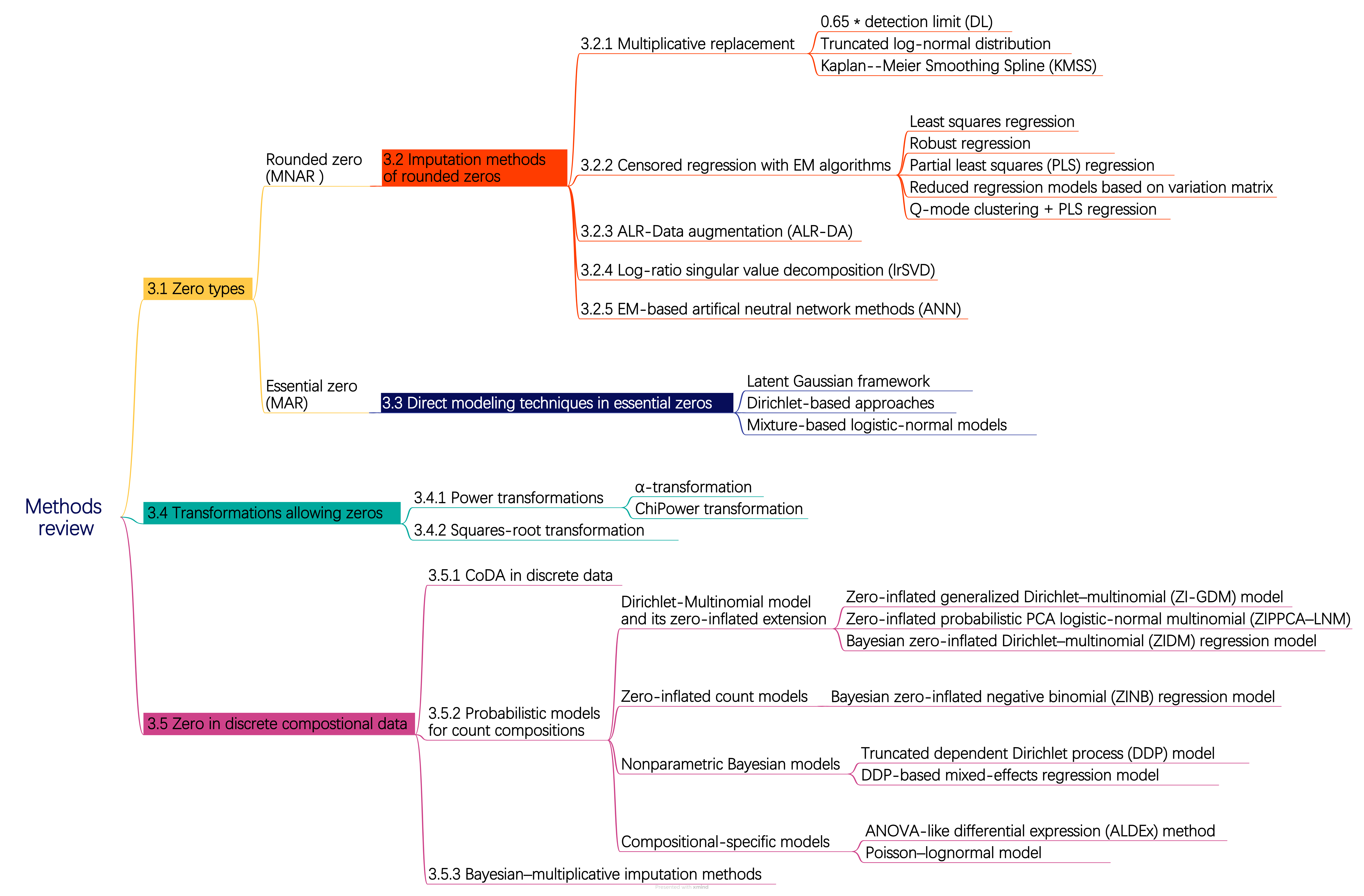}
    \caption{Diagram outlining the structure of Sec.~\ref{sec:rev}. 
    This review covers the classification of zeros, the methods for handling zeros arising from different underlying mechanisms, zero-tolerant transformations, and a dedicated discussion of issues specific to discrete compositional count data. The diagram also indicates the corresponding section numbers associated with each component of the review.}
    \label{fig:review_outline}
\end{figure}

Before reviewing methods for handling zeros in compositional data, we first categorize zeros into two main types. The structure of this section is summarized in Fig.~\ref{fig:review_outline}.

\subsection{Classification and interpretation of zero values}
\label{sec:zero_type}

Distinguishing the nature of zero values is a prerequisite for selecting appropriate handling strategies in compositional data analysis. The first major category is rounded zeros, which arise when a true positive value falls below the detection limit of a measurement procedure and is therefore recorded as zero \citep{palarea2008}. Such zeros may result from technical constraints (e.g., low sequencing depth, instrument sensitivity) or from limited observation windows, during which a present component is not detected \citep{PawlowskyGlahnBuccianti2011}. For example, in household budget surveys, some families may report zero spending on alcohol because no expenditure occurred during the short recording period, even though they do consume alcohol occasionally. These zeros do not reflect true absence but rather under-detection, sampling limitation, or rounding effects \citep{FryChong2005}. From a CoDA perspective, rounded zeros are treated as latent small positive values that require replacement before log-ratio transformations can be applied. As introduced in Sec.~\ref{sec:rounding}, zero-replacement (imputation) methods substitute zeros with small positive constants (or model-based estimates) to reconstruct a strictly positive dataset allowing log-ratio methods to be applied.

A fundamentally different category is essential zeros (also called structural zeros) \citep{Aitchison1986}. These represent the true absence of a component and arise from inherent structural, biological, or chemical constraints—such as chemical incompatibility, biological impossibility, or deliberate exclusion during sampling. Unlike rounded zeros, essential zeros are meaningful: they express that the abundance of a component is genuinely zero. In the context of CoDA, the presence of essential zeros has profound implications. Since all information in compositions resides in ratios \citep{AitchisonKay2003}, a component with an essential zero does not contribute relative information to the composition. An observation containing an essential zero lies on the boundary of the simplex, reducing its effective dimensionality \citep{Bear2016}. Thus, essential zeros are not merely problematic for computation—they change the geometry of the space in which the data reside. Therefore, methods designed for rounded zeros are not applicable. Replacing essential zeros with small positive values, as done in imputation-based approaches, would artificially introduce information where none exists, distorting the statistical meaning of the composition and leading to misleading inference. At the same time, directly applying log-ratio transformations is infeasible, since logarithms of zero are undefined and replacement would violate the principles of logarithmic coherence and scale invariance \citep{FilzmoserHronTempl2018}. For this reason, essential zeros require explicit statistical modeling, such as mixture models, latent variable approaches, or zero-adjusted distributions (see Sec.~\ref{sec:essential}), rather than zero replacement.

Next, we connect the classification of zeros in CoDA with missing-data mechanisms. 
From a statistical perspective, missingness is typically described by how it depends 
on the observed and unobserved parts of the data. Consider complete data 
$\mathbf{X} = (\mathbf{X}_{\mathrm{obs}}, \mathbf{X}_{\mathrm{mis}})$, where 
$\mathbf{X}_{\mathrm{obs}}$ contains the observed values and $\mathbf{X}_{\mathrm{mis}}$ 
the unobserved ones. Missingness is said to be 
Missing Completely At Random (MCAR) when the chance that an entry is missing 
does not depend on either $\mathbf{X}_{\mathrm{obs}}$ or $\mathbf{X}_{\mathrm{mis}}$. 
A weaker and more realistic assumption is 
Missing At Random (MAR), in which missingness may depend on the observed 
values $\mathbf{X}_{\mathrm{obs}}$ but not on the unobserved part 
$\mathbf{X}_{\mathrm{mis}}$. From a practical standpoint, structural zeros can often 
be handled in analogy to MAR-type mechanisms, although their substantive 
interpretation differs substantially \citep{boogaart2013analyzing}.

When the probability that an entry is missing depends on the unobserved values 
$\mathbf{X}_{\mathrm{mis}}$, the data fall under 
Missing Not At Random (MNAR). MNAR mechanisms are the most challenging to 
address because they require explicit knowledge of why the values are missing 
\citep{boogaart2013analyzing}. In compositional data analysis, rounded zeros provide 
a canonical example: they occur when the true value falls below the detection limit 
and is recorded as zero. Rounded zeros can therefore be viewed as left-censored 
observations with a known threshold and are typically treated as MNAR 
\citep{boogaart2013analyzing}.

\subsection{Imputation methods of rounded zeros}
\label{sec:rounding}
Traditional general-purpose imputation methods—such as random forest-based 
imputation (\texttt{missForest}) \citep{stekhoven2011}, predictive mean matching 
(\texttt{mice}) \citep{vanbuuren2011}, and $k$-nearest neighbor imputation 
(\texttt{VIM}) \citep{kowarik2016}—are not well suited for compositional data 
containing rounded zeros. These zeros are effectively left-censored values, yet 
such methods fail to simultaneously account for censoring and the geometric 
constraints inherent to compositional data (see Sec.~\ref{sec:composition}). 
Through comparative analysis, \citet{Templ2021} demonstrated that traditional 
imputation methods often struggle to handle such data effectively.

In the following, we review specialized zero-replacement methods developed for rounded zeros, which incorporate detection-limit information and maintain the relative structure of compositions.

\subsubsection{Multiplicative replacement}
The multiplicative replacement approach handles zeros by first inserting a small 
positive value $\delta$ into every zero component and then rescaling the remaining 
non-zero components so that the composition remains closed. In other words, the 
procedure consists of two coordinated steps: (i) replace each zero $x_{ij}=0$ with 
a small imputed value $\delta_{ij}$, and (ii) proportionally adjust all non-zero 
parts to compensate for the added mass. This guarantees that the total remains 
unchanged and that the resulting composition is compatible with the algebra of the 
simplex \citep{MartinFernandez2003}. Formally, the replacement is defined by
\begin{equation}
r_{ij} =
\begin{cases}
\delta_{ij}, & \text{if } x_{ij} = 0, \\[1ex]
x_{ij}\!\left( 1 - \sum_{k\,:\,x_{ik}=0} \frac{\delta_{ik}}{C_i} \right), & \text{if } x_{ij} > 0,
\end{cases}
\label{eq:multiplicative_replacement}
\end{equation}
where $x_{ij}$ denotes the original composition, and $r_{ij}$ the corresponding 
value after replacement in sample $i$ and component $j$. The term $\delta_{ij}$ is 
the small positive value inserted in place of the zero, and 
$C_i = \sum_{j=1}^{D} x_{ij}$ denotes the original sample total, ensuring that 
$\mathbf{r}_i \in \mathcal{S}^D$. This method provides a simple yet coherent way to 
handle zeros: it preserves closure and subcompositional coherence and remains fully 
compatible with Aitchison geometry on the simplex.
The value $\delta_{ij}$ can in principle be chosen in many ways.
For example, one option is to draw nondetected values from a uniform distribution on $[0,\mathrm{DL}_j]$ \citep{Dorey1993}, although this approach is not implemented in the R package \texttt{zCompositions} \citep{PALAREAALBALADEJO201585}.
In practice, the R package \texttt{zCompositions} provides three specific multiplicative replacement strategies, which we describe below.

A commonly recommended choice for $\delta$ is a constant fraction of the detection limit (DL), typically between 50\% and 70\% \citep{MartinFernandez2003}. It is important to note that, in practical applications, each part of the composition may have a different detection limit, and these limits can even vary across observations. When the proportion of non-detected components is small (e.g. $<10\%$), $\delta_{ij} = 0.65\,\mathrm{DL}_j
$ has been suggested as an optimal value to minimize distortion in the covariance structure \citep{Palarea2014}. The method is simple, computationally efficient, and applicable to high-dimensional data. However, using a component-specific fixed $\delta_{ij}$ means that all zeros within
the same column are replaced by the same value, introducing artificial
similarity among samples for that component. Such artificial structure can
affect covariance relationships and thereby distort multivariate analyses,
particularly when zeros are frequent 
 \citep{martin2012}. This approach is implemented in the \texttt{zCompositions} package as the function \texttt{multRepl()}.


The \texttt{zCompositions} package also provides the function \texttt{multLN()}, 
which assumes that the complete data follow a lognormal distribution. The function 
offers two options: when \texttt{random = FALSE}, censored values are replaced by 
the expected value (geometric mean on the log scale) of the lognormal distribution 
truncated at $\mathrm{DL}_{j}$; when \texttt{random = TRUE}, they are imputed by 
random draws below the detection limit, i.e., from the lognormal distribution 
truncated at $\mathrm{DL}_{j}$ \citep{PALAREAALBALADEJO201585}.In this case, the imputed values $\delta_{ij}$ are generated from the 
component-specific lognormal model truncated at the detection limit, i.e.,
\[
\delta_{ij} \sim \text{LogNormal}_j\!\left(
\hat{\mu}_{\mathrm{LN},j},\, 
\hat{\sigma}_{\mathrm{LN},j}^2
\right)
\quad \text{restricted to } (0, \mathrm{DL}_{ij}).
\]

where $\hat{\mu}_{\mathrm{LN},j}$ and $\hat{\sigma}_{\mathrm{LN},j}^2$ are the 
estimated parameters for component $j$, obtained in log-space using maximum 
likelihood estimation \citep{Palarea2014}. The remaining non-zero components are 
then multiplicatively adjusted according to Equation~\eqref{eq:multiplicative_replacement}. 
This approach is technically sound, computationally efficient, and particularly 
suitable for right-skewed data. \citet{Palarea2014} also introduced a multivariate extension of the lognormal multiplicative replacement, in which non-detected components are imputed using conditional distributions in the ALR space under an additive lognormal (ALN) model, thereby accounting for inter-component covariance rather than imputing components independently from univariate truncated lognormal models. 

A third multiplicative replacement strategy for determining $\delta_{ij}$ is the 
Multiplicative Kaplan--Meier Smoothing Spline (KMSS) method, which is 
implemented in the function \texttt{multKM()} in the R package \texttt{zCompositions}. 
The KMSS approach combines the non-parametric Kaplan--Meier estimator with smoothing 
splines to model the distribution of left-censored values, allowing flexible handling 
of datasets with multiple detection limits or complex censoring structures. The method 
supports both single imputation, where censored observations are replaced 
using their geometric mean, and multiple imputation, achieved by specifying 
the number of simulated points (\texttt{n.points}) and performing repeated sampling 
runs~\citep{PALAREAALBALADEJO201585}. After imputing left-censored entries, KMSS 
applies a multiplicative adjustment to all non-zero components, as required by 
Equation~\eqref{eq:multiplicative_replacement}, to preserve the compositional total. 
Owing to its reliance on non-parametric estimation and spline smoothing, KMSS is 
particularly well suited for sequencing-derived compositional data exhibiting 
multiple censoring thresholds, irregular left-censoring patterns, or heterogeneous 
detection limits.

Although these three methods theoretically ensure that the imputed values $\delta_{ij}$ remain positive and below their corresponding detection limits ($0 < \delta_{ij} < \mathrm{DL}_{j}$), the subsequent multiplicative closure step can occasionally produce numerical artifacts. Specifically, when a given sample $\mathbf{x}_i$ contains a large proportion of censored components or exhibits extremely small observed parts, the rescaling required to restore closure may disproportionately down-weight the observed components. As a result, the adjustment term in Equation~\ref{eq:multiplicative_replacement} can exceed the original magnitude of certain parts, leading to negative reconstructed values for all components in that sample. This phenomenon, though rare, has been noted in previous studies and reflects the instability of multiplicative adjustment under near-degenerate compositions \citep{Palarea2014}.

\subsubsection{Censored regression with Expectation-Maximisation (EM) algorithms}
\label{sec:EM}
The second class of methods belongs to parametric approaches. The modified EM-ALR algorithm, first proposed by \citet{Palarea2007,palarea2008} and later extended with the iterative regression procedure in \citet{HronTemplFilzmoser2010}, assumes that the ALR-transformed data follow a multivariate normal distribution and performs censored regressions between components containing zeros and the remaining components.
Let $\mathbf{X}_{\mathrm{imp}}$ denote the compositional data matrix in which all zeros have been replaced by starting or previously imputed values, and let $\mathbf{Z}_{\mathrm{ALR}} = \text{ALR}(\mathbf{X}_{\mathrm{imp}})$ be its ALR-transformed counterpart. For a column $\mathbf{z}_{\mathrm{ALR},j}$ of $\mathbf{Z}_{\mathrm{ALR}}$ containing 
censored values (corresponding to zeros in $\mathbf{X}$), a regression model is built 
using the remaining columns $\mathbf{Z}_{\mathrm{ALR},-j}$ as predictors:
\[
\mathbf{z}_{\mathrm{ALR},j}
    = \mathbf{Z}_{\mathrm{ALR},-j}^{\prime} \boldsymbol{\beta} + \varepsilon.
\]

At the $l$-th iteration of the EM procedure, the unobserved values 
$z_{\mathrm{ALR},ij}^{(l)}$ are imputed by their conditional expectations under a 
truncated normal distribution that incorporates information from both the observed 
variables $\mathbf{Z}_{\mathrm{ALR},-j}$ and the detection limits $\mathrm{DL}_j$:
\[
\hat{z}_{\mathrm{ALR},ij}^{(l)}
    = \mathbf{Z}_{\mathrm{ALR},i,-j}^{(l)\prime} 
      \hat{\boldsymbol{\beta}}_{\mathrm{LS}}^{(l)}
      - \hat{\sigma}^{(l)}
      \frac{
        \phi\!\left(
            \frac{\psi_{ij} 
            - \mathbf{Z}_{\mathrm{ALR},i,-j}^{(l)\prime}
              \hat{\boldsymbol{\beta}}_{\mathrm{LS}}^{(l)}
            }{
            \hat{\sigma}^{(l)}
            }
        \right)
      }{
        \Phi\!\left(
            \frac{\psi_{ij} 
            - \mathbf{Z}_{\mathrm{ALR},i,-j}^{(l)\prime}
              \hat{\boldsymbol{\beta}}_{\mathrm{LS}}^{(l)}
            }{
            \hat{\sigma}^{(l)}
            }
        \right)
      },
\]
where $\hat{\sigma}^{(l)}$ is the estimated conditional standard deviation of 
$z_{\mathrm{ALR},j}^{(l)}$, and $\hat{\boldsymbol{\beta}}_{\mathrm{LS}}^{(l)}$ is the 
least-squares regression coefficient vector. The term 
$\psi_{ij} = \ln\!\left( \frac{\mathrm{DL}_{j}}{x_{\mathrm{ref}}} \right)$ denotes the 
ALR-transformed detection limit. $\phi(\cdot)$ and $\Phi(\cdot)$ are the standard 
normal density and cumulative functions. After convergence, the imputed ALR data 
$\hat{\mathbf{Z}}_{\mathrm{ALR}}$ are transformed back to the simplex space using the 
inverse ALR transformation.

Compared with non-parametric multiplicative replacement, these modified EM approaches better preserve covariance structure and maintain simplex constraints inherent to log-ratio methodology~\citep{Palarea2007}. Moreover, by incorporating detection limits, they avoid potential failures of traditional EM algorithms under MNAR~\citep{Palarea2007}. Starting values for non-detects $x_{ij}$ can be set to 65\% of the detection limit~\citep{Palarea2007,palarea2008,martin2012} or obtained via $k$-nearest neighbor (KNN) imputation~\citep{HronTemplFilzmoser2010}. Although the ALR transformation is not isometric, \citet{palarea2008} demonstrated 
that the EM algorithm with least-squares regression is invariant under permutations 
of the compositional parts. In practice, this means that changing the ALR reference 
component does not affect the resulting imputations.

Building on the ALR-based EM algorithm described above, 
\citet{martin2012} proposed a robust extension that replaces least-squares 
regression with an affine–equivariant MM-estimator in order to improve 
stability under outliers and small values.
For robust methods, which are affine equivariant and thus compatible with the ILR transformation, the ALR transformation is unsuitable because it is neither permutation-invariant nor isometric. Therefore, the ILR transformation ($\mathbf{Z}_{\mathrm{ILR}} = \text{ILR}(\mathbf{X})$) is more appropriate for robustification and provides a more reasonable treatment of small values and outliers \citep{martin2012}.

The regression coefficients at the $l$-th iteration can be obtained as
\[
\hat{\boldsymbol{\beta}}_{(\mathbf{MM})}^{(l)} 
= \arg\min_{\boldsymbol{\beta}} 
\sum_{i=1}^{n} 
\rho\!\left(
\frac{
    z_{\mathrm{ILR},ij}^{(l)}
    - \mathbf{Z}_{\mathrm{ILR},i,-j}^{(l)\prime} 
      \boldsymbol{\beta}
}{
    \hat{\sigma}^{(l)}
}
\right),
\]
and the update step as
\[
\hat{z}_{\mathrm{ILR},ij}^{(l)} 
= \mathbf{Z}_{\mathrm{ILR},i,-j}^{(l)\prime} 
  \hat{\boldsymbol{\beta}}_{\mathrm{MM}}^{(l)}
- \hat{\sigma}^{(l)} 
  \frac{
    \phi\!\left(
        \frac{
            \psi_{ij}
            - \mathbf{Z}_{\mathrm{ILR},i,-j}^{(l)\prime}
              \hat{\boldsymbol{\beta}}_{\mathrm{MM}}^{(l)}
        }{
            \hat{\sigma}^{(l)}
        }
    \right)
  }{
    \Phi\!\left(
        \frac{
            \psi_{ij}
            - \mathbf{Z}_{\mathrm{ILR},i,-j}^{(l)\prime}
              \hat{\boldsymbol{\beta}}_{\mathrm{MM}}^{(l)}
        }{
            \hat{\sigma}^{(l)}
        }
    \right)
  }.
\]
To address situations where the number of variables far exceeds the number of 
observations—rendering both least-squares and robust regression unstable—
\citet{TemplHronFilzmoseGardlo2016} introduced a partial least squares (PLS)-based regression 
framework for imputing censored ILR coordinates, where
\[
z_{\mathrm{ILR},j}^{(l)} 
= \mathbf{Z}_{\mathrm{ILR},-j}^{(l)\,\prime} \boldsymbol{\beta} + \varepsilon,
\qquad
\mathbf{Z}_{\mathrm{ILR},-j}^{(l)} = \mathbf{T}\mathbf{P}^{\prime}.
\]
Here, $\mathbf{T}$ represents the latent scores and $\mathbf{P}$ the loadings,
with the optimal number of components $k \ll D$ selected by bootstrap.
The ILR transformation is still applied in this case.

\citet{TemplHronFilzmoseGardlo2016} also proposed using the variation matrix
to identify components strongly associated with the target and construct reduced
regression models:
\[
t_{jk} 
= \operatorname{var}\!\left\{
    \ln\!\left(\frac{x_{ij}}{x_{ik}}\right)
  \right\}, 
\quad
i = 1,\dots,n,\; j,k = 1,\dots,D.
\]
Low $t_{jk}$ values indicate strong proportionality \citep{Aitchison1986}, so the components with the smallest $t_{jk}$ values relative to the target 
part are selected as predictors in the reduced model. Building on the PLS-based regression framework, \citet{Chen2017} further
introduced a Q-mode clustering strategy in which the variation matrix is used as
a dissimilarity measure to group components exhibiting similar log-ratio
variability. PLS regression is then applied separately within each resulting
subcomposition to impute the rounded zeros more efficiently and in a manner
that better reflects local dependence structures.

These three regression-based methods are implemented in the R package
\texttt{robCompositions} \citep{templ2011robcompositions} through the
\texttt{imputeBDLs()} function with different method options.

Finally, \citet{Hijazi2011} proposed a related approach using beta regression under a Dirichlet‐based modeling framework to handle rounded zeros in compositional data.

\subsubsection{ALR-Data augmentation (ALR-DA)}

\label{sec:DA}
The ALR-DA algorithm is a data augmentation (DA) approach \citep{TannerWong1987} based on the truncated ALN model, implemented within a Bayesian framework. The algorithm employs a Markov Chain Monte Carlo (MCMC) iterative simulation procedure to draw samples from the joint posterior distribution of the nondetected components and the ALN parameters. Specifically, at the $l$-th iteration, nondetects in the ALR-transformed space are imputed as
\begin{equation}
    \hat{z}_{\mathrm{ALR},ij}^{(l)} \sim 
    \text{TruncNormal}_j\!\left( 
        \mathbf{z}_{\mathrm{ALR},i,-j}^{(l)\prime} \hat{\boldsymbol{\beta}}_j^{(l)},\;
        \hat{\sigma}_j^{2(l)},\;
        z_{\mathrm{ALR},ij} < \mathrm{DL}_{ij}
        \,\middle|\,
        \hat{\boldsymbol{\mu}}^{(l)},\, \hat{\boldsymbol{\Sigma}}^{(l)}
    \right).
\end{equation}

Similar to the ALR-EM algorithm in Sec.~\ref{sec:EM}, the ALN model representation corresponds to regression equations in the ALR space subject to DL constraints, with alternating updates of nondetects and parameters. Unlike ALR-EM, where nondetects and parameter estimates are updated deterministically, ALR-DA performs stochastic updates via simulation. Thus, it can be regarded as a Bayesian alternative to maximum-likelihood estimation for small datasets \citep{PALAREAALBALADEJO201585}, where parameter inference arises from the simulated draws, and covariance-matrix estimates from the imputed data can be computed directly without variance corrections \citep{Palarea2014}. External information can be incorporated through prior specification to improve predictive performance \citep{PALAREAALBALADEJO201585}, although in practice the usual conjugate normal--inverse-Wishart family with noninformative priors is typically assumed. 

Convergence of the DA algorithm often requires additional diagnostic tools, but for the normal family it is generally accepted that approximately 1{,}500 iterations are sufficient to ensure convergence in most applications \citep{Palarea2014}.  

Finally, the imputed compositions can be back-transformed into the original scale using the inverse ALR transformation, as implemented in the \texttt{lrda()} function of the \texttt{zCompositions} package \citep{PALAREAALBALADEJO201585} in \textsf{R}. 

In addition, the \texttt{lrda()} function also provides a multiple-imputation (MI) scheme. The MI approach is particularly effective for multivariate data under the assumption of multivariate normality \citep{Palarea2007}. For each missing value, MI generates multiple plausible values by employing MCMC algorithms to sample from the posterior predictive distribution. These individually imputed datasets are subsequently combined to obtain a single global estimate and its associated variance according to Rubin’s rules \citep{rubin2004multiple}. This strategy has been illustrated, for example, in \citet{martin2003mcmc}, but it is not well suited to datasets with a large proportion of zeros.

As an alternative to MI, \citet{Palarea2014} introduced a bootstrap-based resampling method to propagate the uncertainty of imputation. Their work also details how to integrate censored-data treatment with a bootstrap inference procedure, thus offering a computationally convenient approach for variance estimation when standard MI becomes inefficient.



\subsubsection{Log-ratio singular value decomposition (lrSVD)}
\citet{PalareaAlbaladejoMartinFernandezRuizGazenThomasAgnan2022} proposed the lrSVD algorithm, a method designed for imputing missing values in compositional data. Its core idea is to extract the dominant underlying structure through low-rank matrix approximation in log-ratio space. Specifically, the algorithm employs singular value decomposition (SVD) to find a low-rank matrix $\mathbf{R}$ that best approximates the original data matrix $\mathbf{X}$ (which contains zeros or missing values) in the least-squares sense by minimizing the Frobenius norm: \begin{equation} \min \|\mathbf{X} - \mathbf{R}\|_F^2 \quad \text{subject to} \quad \text{rank}(\mathbf{R}) \leq s, \end{equation} where $s$ denotes the target rank, that is, the number of dominant latent components retained in the approximation. When the data are complete, the optimal solution corresponds to the first $s$ singular values of the SVD of $\mathbf{X}$. However, for incomplete or censored data, the problem is solved iteratively through constrained optimization. To handle rounded zeros, general missing values, or a combination of both, lrSVD first replaces zeros via multiplicative simple replacement using \(0.65 \times \mathrm{DL}\) to obtain an initial intermediate matrix $\mathbf{M}^{(0)}$. This matrix is then transformed into isometric log-ratio (ILR) coordinates (note that this corresponds to the orthonormal log-ratio transformation used in the original paper) and subject to box constraints, leading to the following optimization problem: \begin{equation} \min (1-\beta)\|\mathbf{M} - \mathbf{R}\|_F^2 + \beta\|\mathbf{M} - \mathbf{X}\|_{F_{\text{OBS}}}^2, \quad \text{subject to} \quad \text{rank}(\mathbf{R}) \leq s, \; \mathbf{L} \leq \mathbf{M} \leq \mathbf{U}, \end{equation} where \(0 < \beta < 1\) is a weighting parameter, $\|\cdot\|_F$ denotes the Frobenius norm, and $\|\cdot\|_{F_{\text{OBS}}}$ indicates that the norm is computed only over the observed entries of $\mathbf{X}$, following \citet{PalareaAlbaladejoMartinFernandezRuizGazenThomasAgnan2022}. Here, $\mathbf{M}$ satisfies the box constraints, while $\mathbf{R}$ provides the best rank-$s$ approximation of $\mathbf{M}$. The constraints $\mathbf{L} \leq \mathbf{M} \leq \mathbf{U}$ are applied element-wise; for rounded zeros, $\mathbf{L} = 0$ and $\mathbf{U} = \mathrm{DL}$, whereas for general missing data, $\mathbf{U}$ is set to the column-wise maximum of the observed values. Since the entire procedure is conducted in ILR coordinates, the imputed data preserve essential compositional properties such as scale invariance and subcompositional coherence, without modifying the originally observed values. By properly defining $\mathbf{L}$ and $\mathbf{U}$, the method can simultaneously accommodate rounded zeros, general missing values, and mixed missingness scenarios. The lrSVD algorithm is implemented in the \textsf{R} package \texttt{zCompositions} via the function \texttt{lrSVD()}. However, as an SVD-based approach, lrSVD is sensitive to outliers and the choice of starting values.

\subsubsection{EM-based artificial neural network methods (ANN)}

The method proposed by \citep{Templ2021} integrates a deep artificial neural 
network into an iterative EM framework for rounding-zero imputation.  
The main idea is to exploit the network's ability to learn complex, non-linear
dependencies among compositional parts, while ensuring that the imputed values 
remain strictly below the DL.

As in standard EM procedures, the algorithm begins with an initialization step to 
obtain a complete data set. In the implementation of \citep{Templ2021}, this 
initialization is typically performed using a $k$-nearest-neighbour approach based 
on Aitchison distances and robust means \citep{HronTemplFilzmoser2010}. The E–M updates are 
then carried out in the ILR coordinate system, preserving the fundamental 
principles of compositional data analysis.

Unlike parametric regression-based imputation schemes discussed before, the ANN component does not 
require specification of a model form. Instead, multiple hidden layers and nonlinear 
activation functions allow the network to automatically learn latent interactions 
and complex dependence structures among variables. A drawback of this approach is 
its substantial computational cost, as well as its sensitivity to outliers, which 
can limit robustness in practice.

\subsection{Direct modeling techniques in essential zeros}
\label{sec:essential}

A simple approach to handle essential zeros is amalgamation \citep{Aitchison1986}, which merges components containing essential zeros with related nonzero parts. For example, in household budget data, “alcohol” and “tobacco” may be combined into a single category. However, this comes at a cost: amalgamation sacrifices information about the original parts and is nonlinear in the simplex, limiting interpretability \citep{FilzmoserHronTempl2018}. Since essential zeros carry meaningful structural information, an alternative strategy is to incorporate them directly into the modeling process rather than replace or aggregate them. This allows the intrinsic sparsity of the data to be preserved. Based on this idea, several statistical models have been proposed to handle essential zeros, including mixture-based logistic-normal models, latent Gaussian formulations, and Dirichlet-based approaches.

\citet{AitchisonKay2003} proposed an extension of the logistic normal distribution based on a mixture modeling framework, in which compositional data are partitioned into groups according to their zero patterns, and each group is modeled separately. This approach is particularly suited for continuous or percentage data in which zeros represent true absences. Building on this idea, \citet{Stewart2011} introduced a multiplicative logistic normal mixture model that separately models zero components and defines conditional distributions for the nonzero parts. When a component is nonzero, its log-ratio is assumed to follow a (potentially skewed) logistic normal distribution, increasing flexibility compared to symmetric models. This framework performs well in regression settings and reduces bias introduced by zeros; however, it cannot fully capture covariance among all compositional components, leaving some dependence structures only partially modeled. \citet{Bear2016} further extended this class of models by projecting compositions into lower-dimensional subspaces according to zero patterns and fitting additive logistic normal mixtures with shared location and dispersion parameters across groups. While computationally convenient, this approach does not guarantee subcompositional coherence when zeros are present and assumes identical mean–covariance structures across patterns, which may not hold in heterogeneous datasets. Overall, mixture-based models for essential zeros follow a common principle: separating “zero” and “non-zero” cases, then modeling the positive components via (extensions of) the logistic normal distribution rather than replacing zeros, thus preserving their structural meaning.

The second class of modeling strategies treats zeros as latent variables within a latent Gaussian framework. The basic idea is to first assume a Gaussian distribution in real space and then apply a suitable transformation to map it onto the simplex \citep{ButlerGlasbey2008, Leininger2013, Tsagris2022}. \cite{ButlerGlasbey2008} introduced Euclidean projections onto the simplex, modeling point probabilities on the boundaries through a latent multivariate normal distribution to accommodate zeros. However, their method is restricted to low-dimensional cases and does not preserve subcompositional ratios. Building on this idea, \cite{Tsagris2022} proposed an alternative projection approach for modeling structural zeros based on the multivariate normal distribution, but the zero-censored version they described only allows for one zero per compositional vector, making it effective only when the zero structure is relatively simple. In the context of land use/land cover data, \cite{Leininger2013} advanced the latent variable framework by introducing a power scaling transformation, which maps compositional data into a latent Gaussian space while allowing a local point mass at zero. Within this latent space, spatial dependence can be incorporated via a hierarchical power model, making the approach suitable for higher-dimensional spatial compositional data with essential zeros comparing with \cite{ButlerGlasbey2008}. A key limitation of this approach, however, is that it requires at least one component to remain strictly positive across the spatial domain; otherwise, the transformation is not well-defined.

The third type of model is derived from the Dirichlet distribution. Since compositional data are subject to a fixed sum constraint, the Dirichlet distribution can naturally model the proportions of components while ensuring that they sum to one. \cite{Tsagris2018} proposed a two-layer model called Zero-Adjusted Dirichlet Regression (ZADR). In this approach, observations are first grouped according to their “zero patterns,” and within each group, a truncated Dirichlet distribution is used to model the non-zero components. Dirichlet regression—especially ZADR—provides a natural modeling framework for zero values in compositional data. It requires that all zero patterns share the same set of parameters and is limited by negative correlations and geometric expressiveness. Compared to other methods, it is more practical, although not theoretically the most flexible. \citet{TangWuYangTian2022} proposed a new stochastic representation and developed the Dirichlet Composition Distribution (DCD) as a generalization of the Dirichlet distribution. The DCD separates the modeling of the probability of zero components from the magnitude of nonzero components, effectively handling essential zeros in compositional data. A limitation of the model is that regression modeling with high-dimensional covariates remains challenging and needs further study.

It is important to note that all of the above modeling approaches assume that compositional data are continuous and measured on a real scale. However, sequencing-based microbiome data are fundamentally different: they consist of discrete counts that lie on a lattice rather than a continuum. These discrete and sparse characteristics require specialized treatment beyond continuous CoDA models. Methods developed specifically for count-based compositional data are reviewed in detail in Sec.~\ref{sec:count}.

\subsection{Transformations allowing zeros}
\label{sec:trans0}

In addition to the log-ratio transformations that are incompatible with zeros (see Sec.~\ref{sec: trans}), several alternative transformation approaches have been developed that can directly accommodate compositional data containing zero values.

\subsubsection{Power-based transformations for compositional data}

In compositional data analysis, two traditional approaches are commonly contrasted.
One is the Log-Ratio Analysis (LRA), based on transformations such as $\mathrm{CLR}$ and $\mathrm{ILR}$ (see Section~\ref{sec: trans}), rooted in Aitchison’s geometry \citep{Aitchison1986, FilzmoserHronTempl2018}.
The other is to treat the components as ordinary Euclidean variables, namely Raw Data Analysis (RDA).
While RDA ignores the compositional constraint, it remains simple and widely used; LRA, by contrast, aligns with the CoDA axioms but cannot be applied directly when zeros are present.

To address this issue, \citet{Tsagris2011} proposed the $\alpha$-transformation based on a Box--Cox type power transformation,
which unifies RDA and LRA within a parametric framework and provides a continuum between strictly Euclidean treatments and log-ratio geometry,
offering a practical compromise that enhances feasibility for real-world data—particularly zero-inflated and high-dimensional sparse compositions—while relaxing some of the strict axioms of Aitchison geometry.
The procedure consists of two steps. The first applies the $\alpha$-power transformation to the compositional vector $\mathbf{x}$
followed by a closure operation:
\begin{equation}
u_\alpha(\mathbf{x}) =
\left( \frac{x_1^\alpha}{\sum_{j=1}^D x_j^\alpha}, \ldots,
\frac{x_D^\alpha}{\sum_{j=1}^D x_j^\alpha} \right),
\qquad 0 \leq \alpha \leq 1.
\end{equation}
The second step linearly maps the closed vector into Euclidean space, typically by centering and projecting onto a $(D-1)$-dimensional subspace orthogonal to the unit vector.
This ensures that the transformed data possess Euclidean geometry and can be analyzed using conventional statistical methods.

It is evident that the $\alpha$-transformation includes RDA and LRA as special cases: 
they represent the two endpoints of the parameter space. Adjusting $\alpha$ allows one 
to find compromises—or even superior alternatives—between them. When $\alpha = 1$, 
$u_\alpha(\mathbf{x})$ reduces to a linear scaling of the original components, 
corresponding to RDA. When $\alpha \to 0$, the transformation converges to
\begin{equation}
\ln\left(\frac{x_i}{g(\mathbf{x})}\right), 
\qquad 
g(\mathbf{x}) = \left(\prod_{j=1}^D x_j\right)^{1/D},
\end{equation}
which is the $\mathrm{CLR}/\mathrm{ILR}$ transformation, corresponding to LRA. 
For $0 < \alpha < 1$, the transformation yields a family of geometries lying between 
RDA and LRA. \citet{Tsagris2015} applied the $\alpha$-transformation in discriminant 
analysis and showed that it is more robust than the direct log-ratio transformation in 
the presence of zeros or near-zero-boundary data. \citet{Dong2025Compositional} further 
introduced the $\alpha$-transformation into the Lee Carter-CoDA (LC-CoDA) framework for mortality 
prediction, addressing the limitation that log-ratio analysis cannot be directly 
applied to zero counts. The $\alpha$-transformation has demonstrated strong robustness 
and tunability in supervised learning tasks, including classification and prediction.

On the other hand, \citet{greenacre2024chipower} proposed the ChiPower transformation,
based on a Box--Cox power transformation combined with $\chi^2$ standardization.
Like the $\alpha$-transformation, it belongs to the family of power transformations,
but its procedure differs: ChiPower introduces an additional $\chi^2$ standardization based on column means after the closed power transformation.
In \citet{greenacre2024chipower}, the ChiPower transformation was shown to be particularly effective for unsupervised analyses,
including correspondence analysis. However, while both transformations offer flexibility for zero-valued compositional data by operating between the extremes of ignoring
the compositional constraint and enforcing a strict log-ratio approach,
this flexibility comes at the cost of violating subcompositional coherence,
and the analytical results can be sensitive to the choice of the transformation parameter.
Specifically, the optimal transformation parameter does not have a fixed choice but is typically optimized in a data-driven manner—for instance, by maximizing model likelihood, minimizing cross-validation error, or selecting the value that yields the best model fit or predictive performance—depending on the analytical framework or model to be applied.

\subsubsection{Hyperspherical (directional) transformations for compositional data}

\citet{ScealyWelsh2011} and \citet{SchwobHootenCalzada2024} proposed the use of a square-root transformation 
to map compositional data onto directional data lying on a $(D-1)$-dimensional hypersphere, 
which can then be modeled using directional distributions such as the Kent distribution \citep{ScealyWelsh2011} 
or the elliptically symmetric angular Gaussian (ESAG) distribution \citep{SchwobHootenCalzada2024}. 
These distributions possess flexible covariance structures, allowing for both positive and negative correlations between components, 
thereby overcoming the limitation of the Dirichlet model, which cannot directly model positive correlations. 
However, it should be noted that the square-root transformation does not preserve Aitchison geometry, 
as it fails to maintain the ratios of parts in subcompositions. 
Moreover, when components are close to zero, boundary issues arise in the fitted directional models 
(e.g., the Kent distribution), which may assign non-negligible probability mass outside the positive orthant \citep{SchwobHootenCalzada2024}.


\subsection{Count zeros}
\label{sec:count}

In the following sections, we will focus on methods for handling discrete compositional data. Among such data, those generated by high-throughput sequencing technologies (mentioned in Sec.~\ref{sec:intro}) represent a typical example of discrete compositions, as they are expressed in the form of mRNA transcript counts, OTU counts, and other integer-valued measurements. 

\begin{figure}[t]
 \centering

\includegraphics[width=\textwidth]{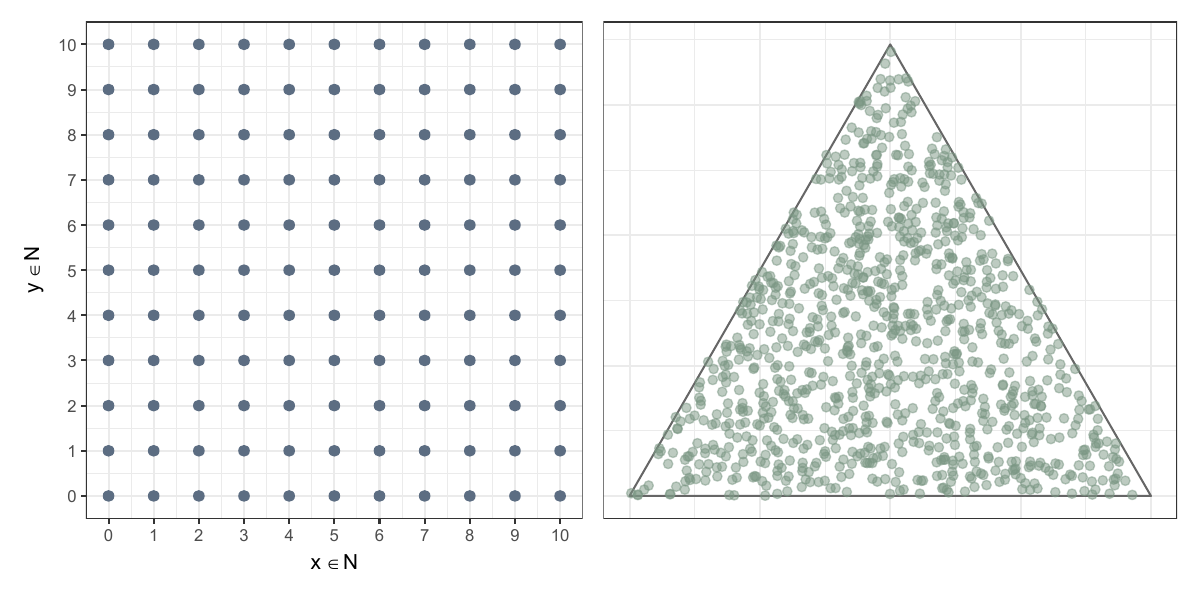}
 \caption{
 Integer lattice $\mathbb{N}^2$ (left) and the continuous 3-part simplex $\mathcal{S}^3$ (right).
Left panel: Shows all $11 \times 11 = 121$ possible integer pairs $(x, y)$ where $x, y \in \{0, \dots, 10\}$.
Right panel: Displays $N=1000$ simulated compositions generated from a symmetric Dirichlet distribution~$\text{Dir}(1, 1, 1)$, projected onto the $2$-simplex.
Count data from HTS technology are realized on a discrete lattice, whereas classical
CoDA assumes that compositions lie in a continuous simplex that is infinitely dense.
 }
 \label{fig:lattice-simplex}
\end{figure}

Before proceeding, it is important to note that integer-valued count data naturally reside on a mathematical structure known as a lattice ~\citep{LovellChuaMcGrath2020}. A $D$-dimensional natural-number lattice $\mathbb{N}^D$ consists of all points of the form $(x_1,\ldots,x_D)$ with $x_i \in \mathbb{N}$, forming a discrete grid shown in left panel of Fig \ref{fig:lattice-simplex} rather than a continuous space. Count vectors generated by high-throughput sequencing—such as transcript counts or OTU counts—are therefore lattice-valued compositions. This lattice structure underlies many of the properties discussed below, including quantization effects, many-to-one closure mappings, and scale-dependent distortions. A detailed description of lattice-valued data and their geometric implications can be found in~\cite{LovellChuaMcGrath2020}. In contrast to lattice-based count data, classical CoDA theory assumes that compositions are defined in a continuous and strictly positive real space. All possible compositions with $D$ parts form the $(D-1)$–dimensional simplex $\mathcal{S}^D$ introduced in Sec.~\ref{sec:composition}, which is a continuous and infinitely dense domain shown in right panel in Fig \ref{fig:lattice-simplex} in which any two compositions can be connected by infinitely many intermediate points. This idealized representation provides the mathematical foundation for log-ratio operations, distance measures, and other analytical tools in CoDA. By contrast, lattice-valued count compositions occupy only a discrete subset of this simplex after closure, and the resolution of these discrete points depends on the underlying sequencing depth. This mismatch in resolution between discrete count data and the continuous simplex lies at the heart of many of the quantization effects and scale-dependent instabilities discussed below. This contrast between lattice-valued count data and the continuous simplex is illustrated schematically in Figure~\ref{fig:lattice-simplex}, which juxtaposes the integer lattice with its continuous compositional counterpart.

Having established the discrete lattice structure of count data, we now discuss the specific empirical properties of HTS-generated compositions. These datasets exhibit two characteristic features. First, zeros and low-abundance components are particularly prominent due to variability in sequence diversity and sampling depth across experimental conditions \citep{LovellChuaMcGrath2020}. Second, while modern sequencing platforms can produce millions of nucleotide reads, the resulting data reflect not only the underlying biological composition but also substantial technical variation introduced by uneven sequencing depth and efficiency biases \citep{Zhang2024}. Third, The discrete nature of count data, with 1 as the smallest unit of resolution, further leads to extremely wide numerical ranges: low-abundance features are concentrated near very small values or zeros, whereas a small subset of highly abundant features may reach tens of thousands or more, inflating the overall mean. This phenomenon is especially evident in RNA-seq datasets \citep{Zhang2024}. Such discreteness and imbalance present challenges for ratio-based CoDA. While CoDA emphasizes relative rather than absolute information between components, the coexistence of many near-zero values, true zeros, and a few extremely large counts can substantially distort log-ratio relationships. 

Next, we introduce the problems caused by the discreteness of count data in CoDA and the current strategies to address count zeros. As noted earlier, zeros can be classified according to their origin into essential zeros and rounded zeros; in discrete data, essential zeros often reflect genuinely low or absent abundance of certain taxa, whereas rounded zeros more commonly arise from insufficient sequencing depth in individual samples \citep{FilzmoserHronTempl2018}. Broadly, the handling of zeros in discrete data can be divided into two main approaches: one focuses on zero replacement (imputation of zeros in discrete data), while the other directly models the data including zeros, thereby retaining the information carried by the zeros themselves. In the following sections, we will discuss these two approaches and their applications in different analytical contexts.

\subsubsection{CoDA in discrete data}
The log-ratio framework of CoDA relies on expressing information through ratios between components, and the transformations introduced in Section~\ref{sec: trans} are all built upon this log-ratio formulation. Moreover, many core statistical quantities in CoDA—such as the variation matrix—are defined in terms of log-ratios \citep{Aitchison1986}.

However, when compositional data appear as discrete counts, some foundational assumptions of CoDA are violated. As discussed earlier, the lattice-valued nature of count data \citet{LovellChuaMcGrath2020} imposes a fixed numerical resolution and prevents the data from occupying the simplex $S^D$ in a continuous manner, thereby challenging the direct application of CoDA to count-based compositions. As noted by \citet{LovellChuaMcGrath2020}, the discreteness of count data introduces quantization error and sampling variation, both of which can substantially affect the results obtained from LRA. The main issues associated with this discreteness can be summarized as follows.

\begin{figure}[htbp]
    \centering
    
    \begin{subfigure}[t]{0.48\textwidth}
        \centering
        \includegraphics[width=\linewidth]{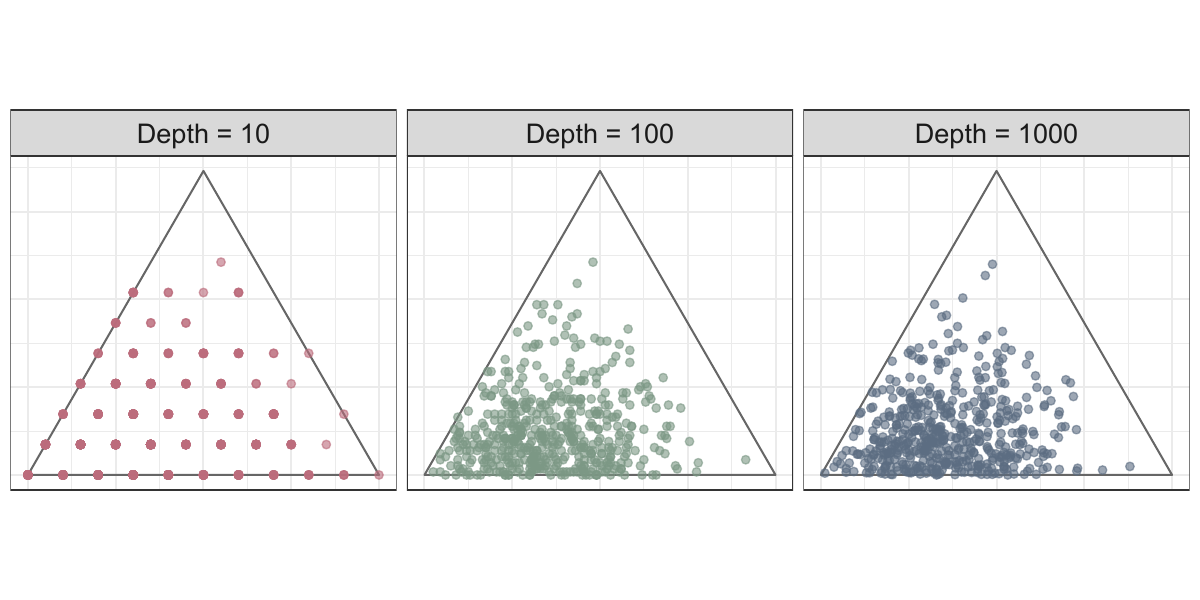}
        \caption{Simplex projection illustrating discretization at different sequencing depths.}
        \label{fig:simplex_depth}
    \end{subfigure}
    \hfill
    \begin{subfigure}[t]{0.48\textwidth}
        \centering
        \includegraphics[width=\linewidth]{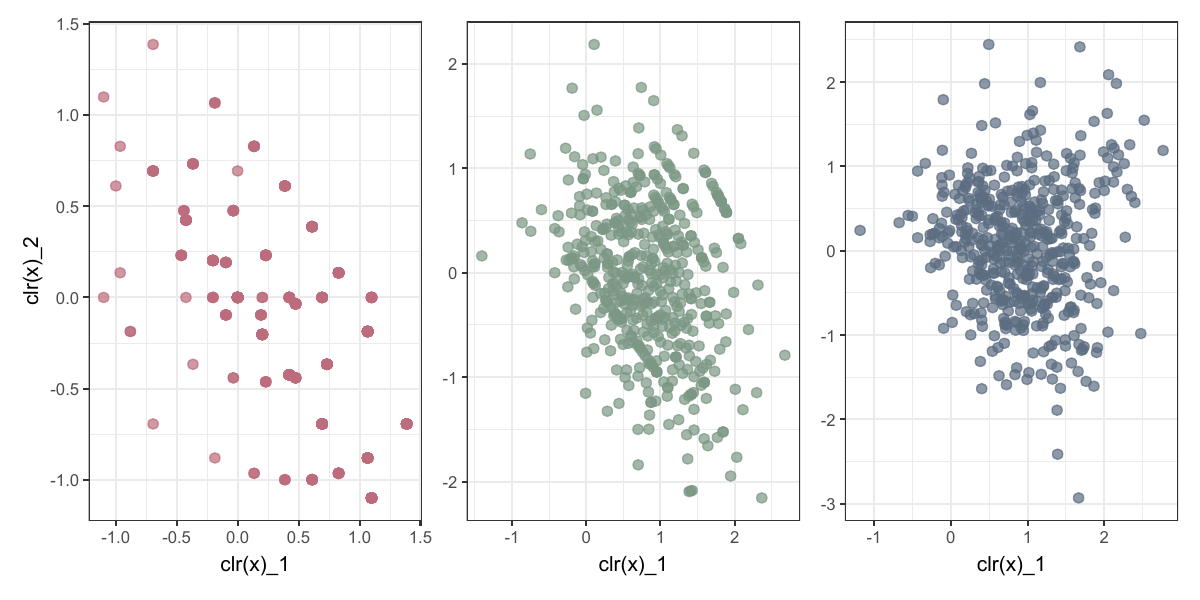}
        \caption{CLR coordinates showing sampling noise and geometric distortion under low depth.}
        \label{fig:clr_depth}
    \end{subfigure}

   \caption{
Comparison of compositional geometry across sequencing depths.
We simulate a sample size of 500 three-part Dirichlet--multinomial compositions with parameters
$\alpha=(4,2,1)$ under sequencing depths $\{10, 100, 1000\}$.
Left (Fig.~\ref{fig:simplex_depth}): shows the resulting compositional points after applying closure and distributes onto the simplex.
Right (Fig.~\ref{fig:clr_depth}): illustrates the corresponding two components of the CLR-transformed composition.
Overall, increasing sequencing depth systematically transitions the compositional data representation from a highly discrete and sparse lattice approximation to a dense, continuous-like geometric structure, which significantly enhances the fidelity and stability of log-ratio analysis.
}
\label{fig:depth_pair}
\end{figure}

A first challenge concerns the closure operation. As previously noted for HTS data, technical limitations lead to substantial variation in sequencing depth across samples, resulting in total counts that differ markedly from one sample to another. To ensure comparability, CoDA typically applies closure, which rescales each composition by its total so that all samples share a common sum and scale invariance is preserved. However, because count data lie on a lattice, closure maps these integer-valued vectors onto only a finite-resolution subset of the simplex. When counts are large, the lattice points become much denser, and their projections onto the simplex more closely approximate the behavior of continuous compositions. In contrast, when counts are small, lattice points are highly sparse, and closure maps them to a coarse and uneven subset of the simplex, making the discreteness of the data particularly pronounced. This scale-dependent change in simplex resolution is illustrated in Fig.~\ref{fig:depth_pair} (left), where shallow sequencing depths yield visibly coarse subsets of the simplex, whereas deeper sequencing produces a much denser and more continuous point cloud. This limited resolution directly contributes to quantization artifacts and instability in downstream log-ratio analyses \citep{LovellChuaMcGrath2020}, as further reflected in the CLR coordinates in Fig.~\ref{fig:depth_pair} (right), where low-depth data exhibit strong discretization and sampling noise.


In addition to the geometric limitations imposed by the lattice structure and closure, discreteness also has direct statistical consequences for log-ratio analysis. The mathematical foundations of LRA are derived under the assumption of strictly positive continuous real numbers \citep{Aitchison1986}, whereas count data are inherently discrete lattice measurements generated through stochastic sampling processes such as multinomial or Poisson distributions \citep{LovellChuaMcGrath2020}. Thus, observed counts represent noisy realizations of underlying biological proportions, and this discreteness becomes especially problematic under shallow sequencing depth. For example, in the same sample, two components with counts of 1 and 2 produce a log-ratio of $\log(2/1)=0.69$, suggesting a twofold difference that largely reflects sampling noise rather than true biological variation. This phenomenon illustrates a broader issue: quantization error is scale dependent \citet{LovellChuaMcGrath2020}. When counts are small, the log-ratio is highly sensitive to single-unit changes, substantially amplifying noise; by contrast, high-count data more closely approximate continuous behavior, making LRA more reliable. As shown in \citet{LovellChuaMcGrath2020}, quantization noise in low-count settings can distort association estimates and obscure true proportional relationships. Consequently, applying LRA directly to low-count data often yields unstable or misleading results.

\begin{figure}[htbp]
    \centering
    
    \begin{subfigure}[t]{0.49\textwidth}
        \centering
        \includegraphics[width=\linewidth]{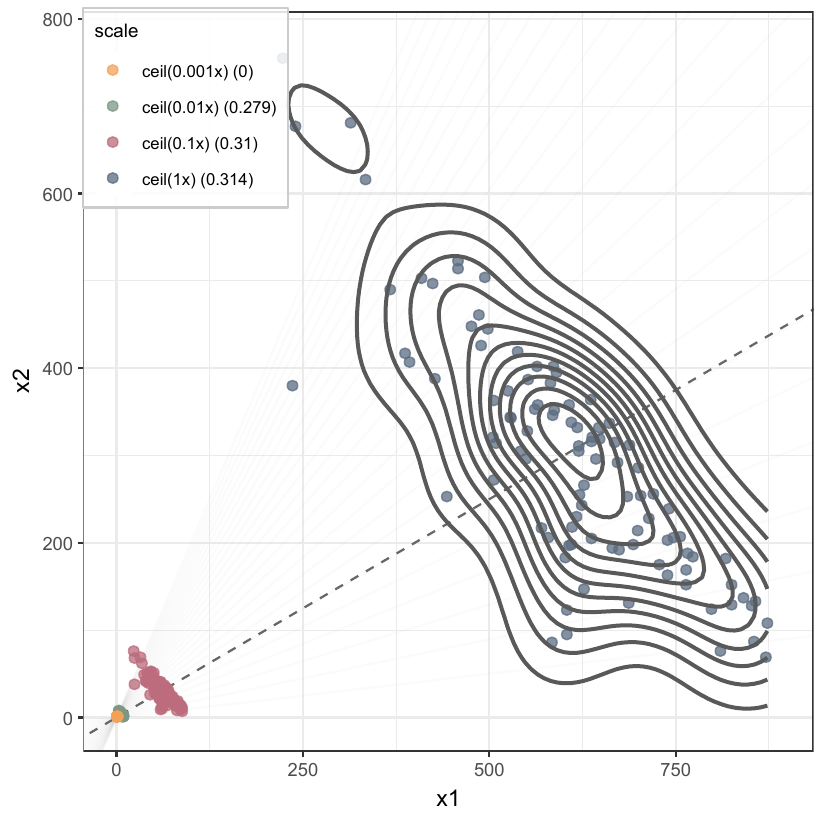}
        \caption{Lattice-valued counts after scaling and ceiling quantization.}
        \label{fig:scaling-ceiling}
    \end{subfigure}
    \hfill
    \begin{subfigure}[t]{0.49\textwidth}
        \centering
        \includegraphics[width=\linewidth]{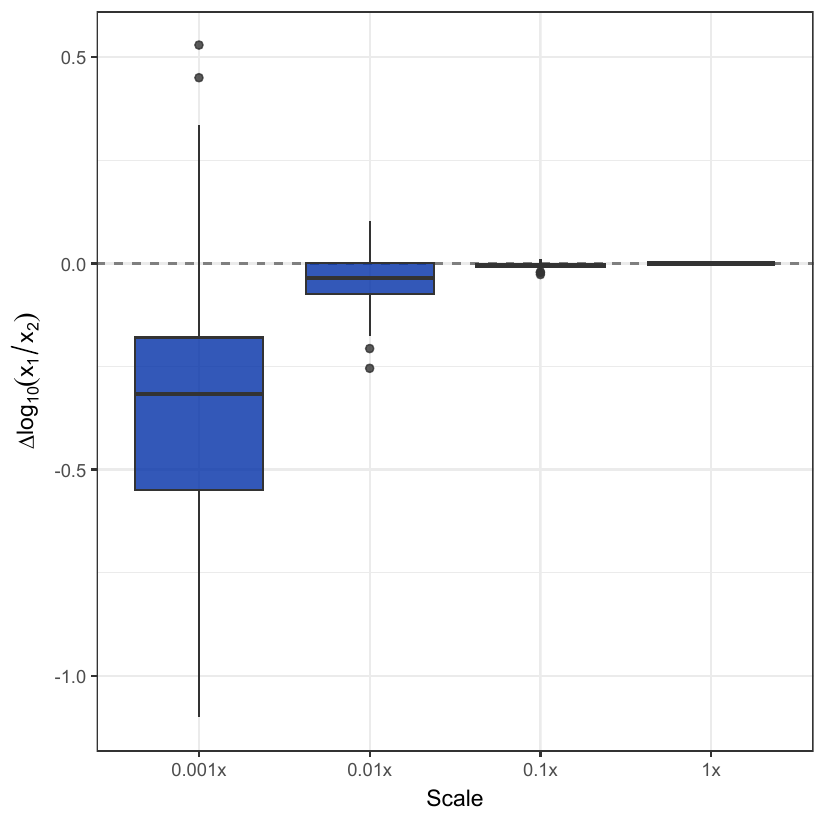}
        \caption{Distribution of per-sample log-ratio shifts.}
        \label{fig:shift-ceiling}
    \end{subfigure}
    
    \caption{
        Effect of global scaling and ceiling quantization on Dirichlet--multinomial count data.
Simulations are conducted using 100 samples from a three-part Dirichlet--multinomial model with sequencing
depth 1000 and $\boldsymbol{\alpha} = (6,3,1)$.
(a): Scatterplots show the quantized lattice-valued pairs $(x_1,x_2)$ obtained after applying four global
scaling factors $s \in \{1, 0.1, 0.01, 0.001\}$. The dashed line indicates the reference proportionality
$x_2 = 0.5x_1$, which corresponds to the expected ratio $E[x_1/x_2] = 2$ under the generative model.
Legend entries report the mean log-ratio $\log_{10}(\overline{x}_1/\overline{x}_2)$ computed at each scale.
(b): Boxplots show the distribution of per-sample log-ratio shifts
$\Delta \log_{10}(x_1/x_2)$ induced by ceiling
    }
    \label{fig:scaling-quantization}
\end{figure}

A core assumption of CoDA is scale invariance: multiplying all components by a constant 
should not alter their relative structure. Count data violate this principle because their 
smallest possible increment is 1. When counts are globally scaled down, many distinct 
compositions collapse onto the same lattice points, creating quantization artifacts 
analogous to reducing the resolution of an image \citep{LovellChuaMcGrath2020}. This 
collapsing of distinct configurations leads to information loss and violates the scale 
invariance assumption fundamental to CoDA. The resulting quantization error is particularly 
severe for low counts, where single-unit changes can substantially distort log-ratios. As 
shown in \citet{LovellChuaMcGrath2020}, small counts can cause pairwise associations to 
fluctuate widely due to sampling noise, making log-ratio–based inference unreliable. To illustrate these effects, we follow an approach analogous to that of 
\citet{LovellChuaMcGrath2020} and demonstrate the impact of quantization in Fig.~\ref{fig:scaling-quantization} by generating
three-part Dirichlet--multinomial counts with a sample size of 100, a sequencing depth of 1000, and $\boldsymbol{\alpha}=(6,3,1)$, and then applying progressively stronger global 
scaling followed by ceiling quantization, thereby revealing how distinct lattice 
configurations merge and how the corresponding log-ratios deviate from their 
unquantized values. The simulated results in Fig.~\ref{fig:scaling-ceiling} 
illustrate this effect: global downscaling followed by ceiling quantization causes 
originally distinct Dirichlet--multinomial compositions to merge onto a limited set 
of lattice locations, demonstrating substantial information loss. The legend reports the mean log-ratio $\log_{10}(\overline{x}_1/\overline{x}_2)$ computed at each scale. As the scaling factor decreases, these mean log-ratios progressively drift away from the true underlying value of $\log_{10}(2)\approx 0.31$, reaching values as low as $0.28$, thereby quantifying the increasing distortion induced by aggressive downscaling and discretization. 
Fig.~\ref{fig:shift-ceiling} shows that this quantization induces 
non-negligible shifts in sample-wise log-ratios. These deviations grow larger as the 
scaling factor decreases, confirming that low-count regimes are especially susceptible 
to quantization-driven distortion of relative information.

In real sequencing data, the distribution of count values is typically extremely wide. Low-abundance features often have only single-digit counts or even zeros, while a few high-abundance features may reach tens of thousands or more. This highly imbalanced distribution reflects both the intrinsic diversity of biological systems and introduces substantial scale differences among components. Therefore, appropriate scaling is often required before analyzing such count compositional data to mitigate the impact of wide-ranging count magnitudes. \cite{Zhang2024} focused on microbiome data as a representative example of count compositional data and emphasized that, due to uneven sequencing depth, samples are not directly comparable unless scaling is applied prior to log-ratio transformation.

In the traditional CoDA framework, discrepancies in total counts across samples are addressed by the closure operation \cite{Aitchison1986}, which transforms absolute information into relative proportions by dividing each component by the sample total. This procedure is conceptually equivalent to the Total Sum Scaling (TSS) approach summarized in \cite{Zhang2024}. While closure restores the compositional meaning of the data mathematically, it may discard biologically important absolute information, such as total species abundance or biomass. On the other hand, \cite{Zhang2024} introduced alternative scaling methods that preserve total-scale information (e.g., variance scaling or z-score normalization). Although these methods can retain some aspects of absolute magnitude, they often distort the original proportional structure, causing the data to no longer satisfy the compositional constraint.

\subsubsection{Probabilistic models for zero-inflated count compositional data}
\label{sec:distributions for count}
The Dirichlet-multinomial (DM) distribution provides a principled framework for modeling compositional count data due to its inherent ability to respect the simplex constraint. However, this model suffers from two key limitations: (1) its inability to handle zero-inflated data, and (2) its restriction to capturing only negative correlations between components. Modern high-throughput sequencing produces microbial abundance data that are both high-dimensionaland extremely sparse, prompting the development of DM extensions that explicitly model the excess zero structure and accommodate more general correlation patterns. A common feature of these extensions is the adoption of a zero-inflated framework, in which zeros are modeled through a two-component mixture consisting of a point mass at zero and a count distribution for the non-zero observations. The associated latent indicators naturally distinguish between structural zeros and rounding zeros, allowing the models to preserve the different meanings carried by zeros in count compositional data. 

\citet{Tang2019} introduced the zero-inflated generalized Dirichlet–multinomial model by modifying the Beta stick-breaking representation of the generalized Dirichlet distribution. Zero inflation is incorporated through zero-inflated Beta variables, enabling more flexible correlation structures than the classical DM. Parameter estimation is performed via an EM algorithm. A limitation of the approach, however, is that it does not naturally extend to multivariate regression settings. \citet{Zeng2023} proposed the Zero-Inflated Probabilistic PCA Logistic-Normal Multinomial (ZIPPCA–LNM) model, which combines a low-rank logistic-normal multinomial formulation with an explicit zero-inflation component. This approach captures complex covariance patterns through a low-rank latent structure while leveraging the flexibility of the logistic-normal distribution. An iterative procedure integrating empirical Bayes estimation and variational approximation enables scalable computation in high-dimensional settings. More recently, \citet{koslovsky2023} introduced a Bayesian zero-inflated Dirichlet–multinomial (ZIDM) regression model that integrates zero inflation, compositional constraints, and variable selection in high dimensions. Zero inflation is modeled via a gamma reparameterization, while sparsity-inducing priors and a dimension-adaptive Metropolis–Hastings sampler improve inference stability and computational efficiency. Taken together, these developments reflect three broad directions in zero-inflated DM modeling: extending distributional flexibility, enriching dependence structures, and integrating zero inflation into high-dimensional regression frameworks. Despite this progress, challenges remain in scalability, interpretability of latent zero mechanisms, and jointly modeling zero inflation, overdispersion, and complex correlation patterns in large-scale microbiome count data. 

Beyond the DM framework, \citet{Jiang2021} proposed a Bayesian zero-inflated negative binomial (ZINB) regression model with spike-and-slab priors for differential abundance analysis. However, because the model does not incorporate compositional constraints or link covariates to the zero-inflation mechanism, its estimates may not adequately account for compositional effects when applied to compositional count data.

In addition to the parametric methods mentioned above, \cite{Ren2017} proposed a more flexible nonparametric approach in model construction that does not require pre-setting a fixed number of parameters. This allows the model complexity to adaptively adjust based on the characteristics of the data, enabling it to capture complex patterns and inherent sparsity. The method employs a truncated dependent Dirichlet process (DDP) prior, which more naturally reflects the correlations between components and the sparsity in the data. Later, \cite{Ren2020} extended this approach to mixed-effects regression models to accommodate the needs of longitudinal data analysis. However, the method does not explicitly model zero-inflation indicators, preventing it from distinguishing between different types of zeros. Additionally, it is designed for small to medium-sized covariate spaces, relying on model comparison techniques for model selection.

Some additional modeling strategies are also worth noting.  ~\cite{Fernandes2013} proposed, within a Bayesian framework, an ANOVA-like analytical approach called ANOVA-Like Differential Expression (ALDEx). This method combines the Dirichlet-multinomial model with ILR-transformation to perform variability assessment and differential expression analysis on sparse compositional data. For compositional data, the counts reflect relative proportions rather than absolute abundances, and the total read count \( n_i \) itself is random and provides limited biological information about the sample. Therefore, ~\cite{Fernandes2013} focused on inferring the species proportions \( p_i \). The approach starts directly from the counts \( n_i \), and employs a Bayesian method combining a multinomial likelihood with a Dirichlet prior distribution (with hyperparameter set to 0.5) to estimate the posterior distribution of \( p_i \). This ensures that even when the observed count for a species is zero, its estimated proportion will not be exactly zero, thus mitigating biases caused by sparsity. Subsequently, the proportions \( p_i \) are transformed using the ILR transformation into linearly independent variables, enabling statistical analysis and variability exploration in Euclidean space. And \cite{BaconShone2008} proposed an approach based on the Poisson-log normal distribution, applicable when essential zeros appear in discrete compositions of count data.

\subsubsection{Bayesian–multiplicative imputation methods}
For HTS-derived compositional data, limitations in sequencing depth often cause zero counts to arise simply from insufficient sampling. Increasing sequencing depth would, in many cases, convert these zeros into small nonzero counts. Therefore, for such sampling zeros, it is reasonable to apply imputation approaches designed for rounded zeros, as discussed in Section~\ref{sec:rounding}.

However, most existing imputation methods assume a continuous data representation and operate within the LRA
framework. This creates a mismatch with count-based compositional data, which are inherently discrete and lack scale
invariance. Under small-count conditions, scaling operations and log-ratio transformations can introduce information
loss and quantization error, distorting the underlying proportional relationships. Consequently, although rounded-zero
imputation provides a practical strategy for addressing sampling zeros, the resulting imputed values are positive
continuous numbers below the detection limit rather than integer counts, altering the discrete structure of the original
data and requiring cautious interpretation.

To address these issues, the following section introduces a Bayesian zero-replacement approach \citep{MartínBayesian, Daunis2008}, 
which integrates Bayesian estimation with classical rounded-zero imputation ideas \citep{MartinFernandez2003}. 
This framework yields posterior predictive proportions that can be mapped back to counts while remaining coherent with the statistical principles of compositional data. 
The method provides a conceptual bridge between CoDA and discrete lattice-valued data and is implemented in the \texttt{zCompositions} package through the function \texttt{cmult()}.

In the context of count compositional data, let $\mathbf{c} = (c_1, \dots, c_k)$ denote the observed counts across $k$ categories, with total $N = \sum_j c_j$. Assuming a multinomial model with parameters $\boldsymbol{\theta} = (\theta_1, \dots, \theta_k)$, the likelihood is
\[
P(\mathbf{c} \mid \boldsymbol{\theta}) = \frac{N!}{c_1! \cdots c_k!} \prod_{j=1}^k \theta_j^{c_j}.
\]

A Dirichlet prior with strength $s$ and center $\mathbf{t} = (t_1, \dots, t_k)$ is assigned to $\boldsymbol{\theta}$, i.e.,
\[
\boldsymbol{\theta} \sim \text{Dir}(s\mathbf{t}), \quad t_j > 0, \; \sum_j t_j = 1.
\]
The posterior is then $\boldsymbol{\theta}|\mathbf{p} \sim \text{Dir}(\mathbf{p} + s\mathbf{t})$, where $\mathbf{p} = \mathbf{c}/N$, and the posterior mean is
\[
\mathbb{E}(\theta_j|\mathbf{p}) = \frac{p_j + s t_j}{N + s}.
\]
When $p_j = 0$, the posterior estimate becomes $\frac{s t_j}{N + s}$, yielding positive replacements even for zero counts. 

Table~\ref{tab: priors} summarizes commonly used Dirichlet priors and their corresponding posterior means.

\begin{table}[ht]
\centering
\caption{Dirichlet priors and Bayesian posterior estimators.}
\begin{tabular}{lcc}
\toprule
\textbf{Prior} & \textbf{s} & \textbf{Posterior mean} \\
\midrule
Haldane       & $0$        & $\frac{p_j}{N}$                 \\
Perks         & $1$        & $\frac{p_j + 1/k}{N + k/2}$      \\
Jeffreys      & $\frac{k}{2}$ & $\frac{p_j + 1/2}{N + k/2}$       \\
Bayes–Laplace & $k$        & $\frac{p_j + 1}{N + k}$                \\
\bottomrule
\end{tabular}
\label{tab: priors}
\end{table}

To preserve ratios between non-zero parts, Bayesian–multiplicative (BM) methods combine the posterior estimator with multiplicative adjustment:
\[
p_j^* =
\begin{cases}
\frac{s t_j}{N + s}, & \text{if } p_j = 0, \\[1ex]
p_j \!\left(1 - \frac{s}{N + s} \sum_{p_k = 0} t_k \right), & \text{if } p_j > 0.
\end{cases}
\]

Although this approach provides coherent replacements for zeros, it lacks scale invariance because the outcome depends on $N$. \citet{MartínBayesian} addressed this by introducing a scaled Dirichlet distribution \citep{PawlowskyGlahnBuccianti2011}.

In addition, \citet{LovellChuaMcGrath2020} proposed a preliminary conceptual framework for discrete replacement, in which zero values are substituted with small discrete positive numbers (e.g., 0.1, 0.2, \dots, 0.9) derived from the empirical distribution of non-zero observations. This approach aims to preserve the discrete nature of count data while maintaining approximate statistical consistency with the original dataset.

\section{Comparison for imputation-based zero handling}
\label{sec:comp}

In the preceding sections, we introduced several strategies for handling compositional datasets with zeros, such as transformations allowing zeros and statistical models that directly account for zero inflation. However, these approaches are typically model-dependent. For example, regression-based models are generally constructed around a response variable to capture relationships or make predictions, while zero-tolerant transformations (e.g., the Box--Cox family) require selecting parameters (such as $\alpha$) that are determined within specific modeling frameworks or optimization criteria. In practical applications, however, one often does not know in advance which statistical methods will be used in downstream analysis. This motivates a separate investigation of imputation methods, whose goal is to complete the data matrix in a way that
remains compatible with common downstream analyses—such as PCA, clustering, and regression—without committing to a
specific modeling framework. 

The present study systematically compares multiple imputation methods using real count-based compositional data, i.e., the Rabbit data ~\citep{greenacre2021compositional}, with the goal of evaluating their ability to restore data completeness, preserve compositional structure, and remain numerically stable in high-dimensional sparse settings. These evaluations allow us to identify the conditions under which each imputation method is most appropriate and to assess whether the resulting completed data can reliably support downstream multivariate analyses.
Such an empirical comparison is essential because count-based compositional data possess distinctive characteristics that challenge traditional imputation frameworks. The data are inherently discrete with a wide dynamic range—most features occur at very low counts or zeros, while a small number of highly abundant components may reach thousands. Moreover, these datasets are often high-dimensional, with the number of features far exceeding the number of samples. These properties introduce difficulties in numerical stability, statistical robustness, and computational efficiency. Motivated by these considerations, we designed real-data experiments to comprehensively evaluate the performance of existing imputation strategies for discrete compositional count data. To further substantiate our findings, we additionally conducted an identical set of experiments using simulation data exhibiting the same discrete and compositional properties as the Rabbit dataset. The corresponding results, provided in Appendix ~\ref{appendix.A}, corroborate the patterns observed in the real-data analysis and strengthen the robustness of our conclusions. All code and materials used in this work are available at \url{https://github.com/tangwenq/Comparison-of-zero-handling-method-in-discrete-count}.

\subsection{The rabbit dataset}

Based on the motivations discussed above, we evaluate the performance of different imputation strategies using the rabbit dataset~\citep{greenacre2021compositional}, a real compositional count dataset. This dataset consists of counts of $D = 3937$ microbial genes measured across $n = 89$ rabbit samples, forming an $89 \times 3937$ matrix.

\begin{figure}[ht]
    \centering
    \begin{subfigure}[b]{0.48\textwidth}
        \centering
        \includegraphics[width=\textwidth]{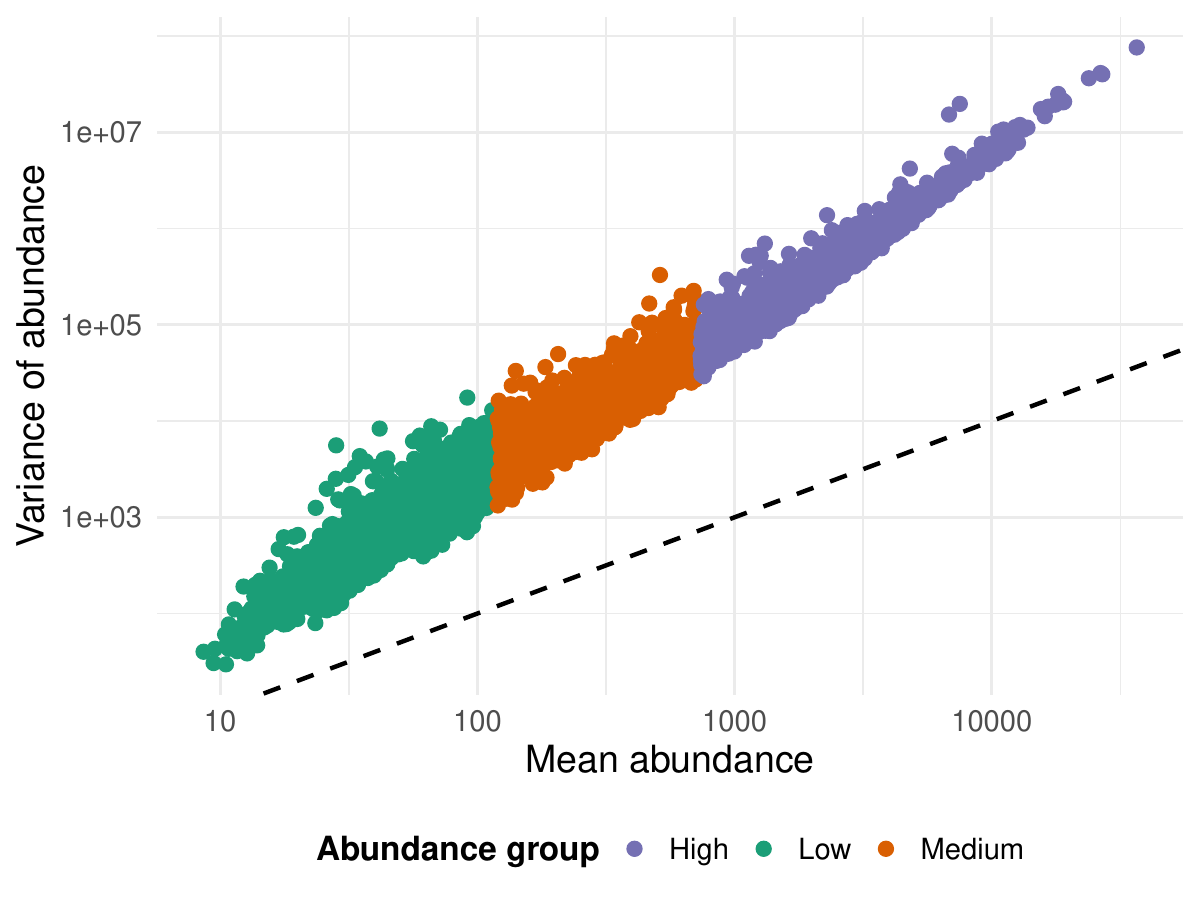}
        \caption{Mean–variance relationship of counts (log–log scale).}
        \label{fig:mean_variance}
    \end{subfigure}
    \hfill
    \begin{subfigure}[b]{0.48\textwidth}
        \centering
        \includegraphics[width=\textwidth]{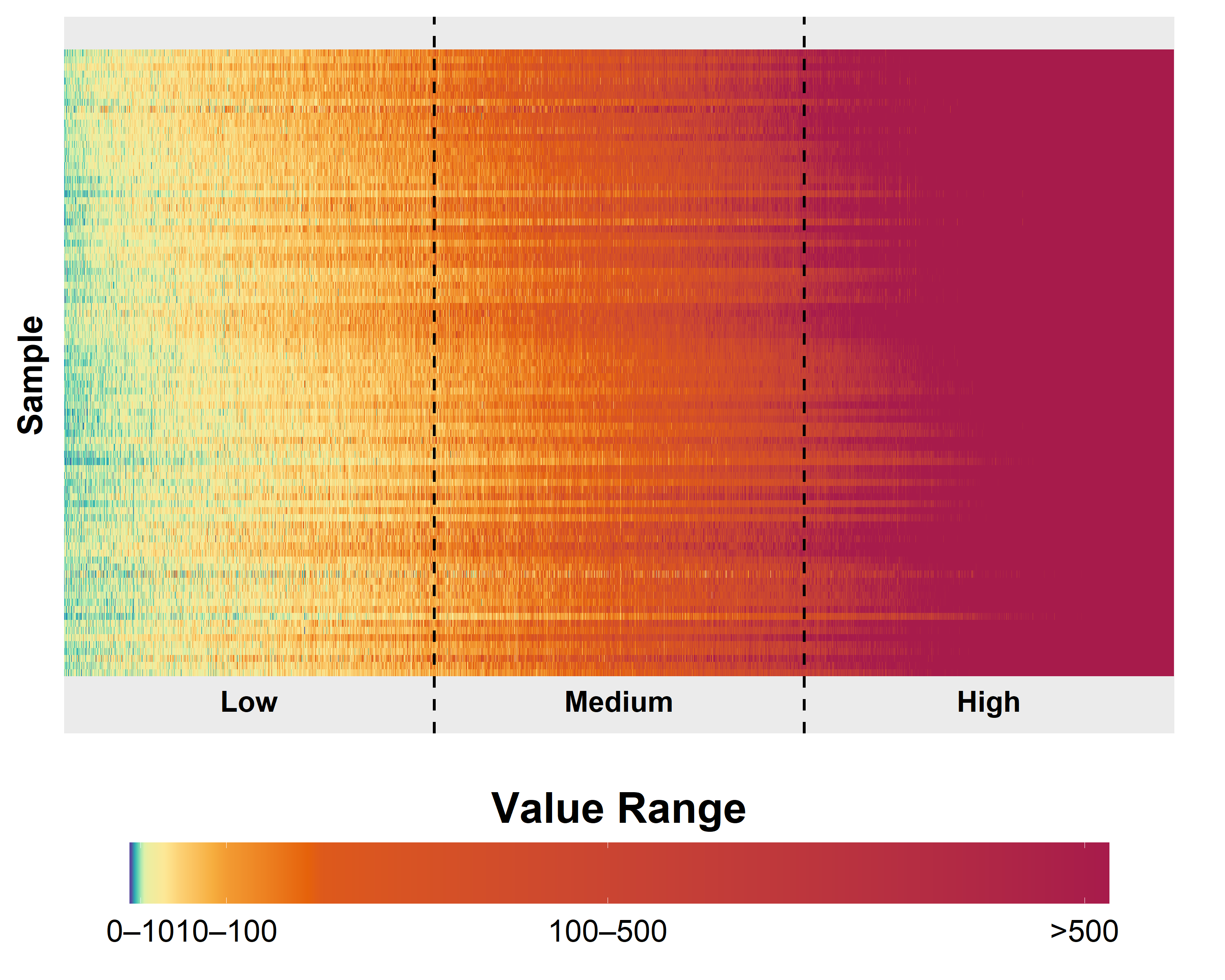}
        \caption{Heatmap of count values.}
        \label{fig:rabbits_heatmap}
    \end{subfigure}

    \caption{
    (\subref{fig:mean_variance}) shows the log–log mean–variance relationship across species,
        illustrating the over-dispersion typical of count-based compositional data.
        (\subref{fig:rabbits_heatmap}) displays the distribution of counts across samples,
        highlighting the wide dynamic range and considerable inter-feature heterogeneity. Low-, medium-, and high-abundance categories were defined by tertiles of the column means.
    }
    \label{fig:rabbit_dataset}
\end{figure}

The rabbit data display key characteristics typical of high-throughput sequencing outputs: wide dynamic range, strong
heterogeneity across features, and clear over-dispersion. As shown in Figure~\ref{fig:mean_variance}, the log--log
mean–variance relationship increases monotonically with the mean, indicating substantial over-dispersion. The heatmap
in Figure~\ref{fig:rabbits_heatmap}, ordered by column means, further illustrates the broad numerical range of counts—
from single-digit values in low-abundance features to counts in the tens of thousands for highly abundant ones. For
visual clarity, low-, medium-, and high-abundance categories in the heatmap are defined by tertiles of the column means.

The dataset is fully observed and contains no zeros, which allows us to introduce zeros at controlled proportions and evaluate imputation performance against known true values. These properties make the rabbit dataset a realistic and suitable benchmark for assessing the stability, robustness, and scale sensitivity of imputation methods for discrete compositional data.

\subsection{Experiment design}
\label{sec:Experiment_design}

In our comparison study, we designed two sets of experiments to systematically examine how both the proportion of zeros and the dimensionality of the data influence the performance of imputation methods.
The first experiment examines how imputation accuracy varies with increasing zero proportions under different dimensional settings. For each chosen dimensionality level ($m = 50, 500, 1000$), we randomly sampled $m$ columns from the Rabbit dataset and introduced rounded zeros by replacing all values below a
column-specific quantile threshold $Q_{p}(x_{\cdot j})
$ with zero, where $p$ denotes the $p$-th quantile level. In our experiments,
$p$ varied from 0.05 to 0.8.
Zeros were inserted into every second column. This procedure produced controlled zero proportions. Each configuration was replicated 500 times. At higher dimensions, however, computational cost and occasional convergence failures of certain imputation algorithms made very large-scale replications infeasible.

The second experiment examined the effect of dimensionality by fixing the zero proportions at $p = 0.2$ and $p = 0.5$
and varying the number of selected columns from $m = 50$ up to $m = 1400$, with 500 replications performed for each
setting. All imputed matrices were obtained using the methods summarized in Table~\ref{sim:imputation table}.
 Because most imputation methods are formulated for continuous data, their outputs are continuous as well. To better adapt these methods to count-valued compositional data and preserve the lattice structure inherent to sequencing counts, we additionally applied an upward rounding step to convert imputed values into integers. This allows us to evaluate whether restoring discreteness affects the performance and numerical stability of each method.

\begin{table}[htbp]
\centering
\caption{Overview of imputation methods for left-censored compositional data}
\label{sim:imputation table}

\begin{tabular}{p{3cm} p{3cm} p{9cm}}
\toprule
\textbf{Method (abbr.)} & \textbf{Function} & \textbf{Description} \\
\midrule
\addlinespace[4pt]

\textbf{mult\_repl} & \texttt{multRepl()} &
Multiplicative replacement using a fixed fraction (0.65) of the DL, i.e.,
$Q_{p}(x_{\cdot j})$.
\\

\addlinespace[4pt]
\textbf{mult\_lognorm} & \texttt{multLN()} &
Lognormal-based replacement using a truncated lognormal model below DL. \\

\addlinespace[4pt]
\textbf{mult\_KMSS} & \texttt{multKM()} &
Multiplicative replacement using a KMSS estimate below DL. \\

\addlinespace[4pt]
\textbf{lr\_da} & \texttt{lrDA()} &
Log-ratio data augmentation procedure using iterative simulation. \\

\addlinespace[4pt]
\textbf{lr\_em} & \texttt{lrEM()} &
EM-based parametric replacement using least-squares regression. \\

\addlinespace[4pt]
\textbf{lr\_SVD} & \texttt{lrSVD()} &
Low-rank log-ratio imputation using singular value decomposition. \\

\addlinespace[4pt]
\textbf{GBM} & \texttt{cmult()} &
Bayesian multiplicative method based on a Dirichlet--multinomial formulation. \\

\addlinespace[6pt]
\midrule
\addlinespace[4pt]

\textbf{lmrob} & \texttt{imputeBDLs(..., method="lmrob")} &
EM-based parametric replacement using robust MM regression. \\

\addlinespace[4pt]
\textbf{PLS} & \texttt{imputeBDLs(..., method="PLS")} &
EM-based parametric replacement using pls regression. \\

\addlinespace[6pt]
\midrule
\addlinespace[4pt]

\textbf{dl\_unif} & \texttt{runif(0.1 * DL, DL)} &
Random draws from a uniform distribution between $0.1 \cdot DL$ and $DL$. \\

\addlinespace[4pt]
\textbf{add1} & -- &
Fixed-value replacement where all zeros are set to 1. \\

\bottomrule
\end{tabular}
\end{table}

\subsection{Distortion assessment}

We use two commonly adopted evaluation metrics in the imputation literature—the Average Difference in Covariance Structure (ADCS) and the Compositional Error Deviation (CED)—to assess how well each method preserves the statistical structure of the data \citep{MartinFernandez2003, PALAREAALBALADEJO201585,templ2011robcompositions}. These metrics quantify, respectively, the change in covariance structure and the deviation in compositional geometry by comparing the imputed matrix with the original complete dataset.

Let $\mathbf{S} = [s_{ij}]$ denote the sample covariance matrix computed from the original data in  ILR coordinates, and let $\mathbf{S}' = [s'_{ij}]$ denote the corresponding covariance matrix computed after imputation. The ADCS is defined as
\begin{equation}
  \mathrm{ADCS} = \frac{1}{D-1} \left\| \mathbf{S} - \mathbf{S}' \right\|_F,
  \label{eq:adcs}
\end{equation}
where $D-1$ is the dimensionality of the ILR-transformed space, and $\|\cdot\|_F$ denotes the Frobenius norm. A value of $\mathrm{ADCS} = 0$ indicates perfect preservation of the covariance matrix after imputation.

Let $n_M$ be the number of compositions $\mathbf{x}_m$ containing at least one 
artificially introduced zero, and $M$ the corresponding index set. 
Let $\tilde{\mathbf{x}}_m$ denote the imputed version of $\mathbf{x}_m$.
Let $\mathbf{X}_{-M}$ denote the subset of fully observed compositions obtained by 
removing all rows indexed by $M$. 
The compositional error deviation (CED) is defined as
\begin{equation}
  \mathrm{CED}
  =
  \frac{1}{n_M}
  \sum_{m \in M}
  \frac{
    d_A(\mathbf{x}_m, \tilde{\mathbf{x}}_m)
  }{
    \max\limits_{\mathbf{x}_i,\,\mathbf{x}_j \in \mathbf{X}_{-M}}
    d_A(\mathbf{x}_i, \mathbf{x}_j)
  },
  \label{eq:ced}
\end{equation}
where $d_A(\cdot,\cdot)$ denotes the Aitchison distance defined in Eq.~\ref{eq:ait_distance}.

The denominator in Equation~\eqref{eq:ced} normalizes the deviation by the maximum observed Aitchison distance between fully observed compositions in the dataset $\mathbf{X}$. This normalization ensures that CED is a relative measure of distortion, accounting for the overall compositional variability of the data. This criterion is a generalization of the measures proposed by \citet{FilzmoserHronTempl2018}.

Both ADCS and CED provide quantitative insights into the extent to which imputation methods preserve essential compositional characteristics and covariance structures of the original data.

\begin{figure}[htbp]
    \centering

    \begin{subfigure}[b]{0.68\textwidth}
        \centering
        \includegraphics[width=\linewidth]{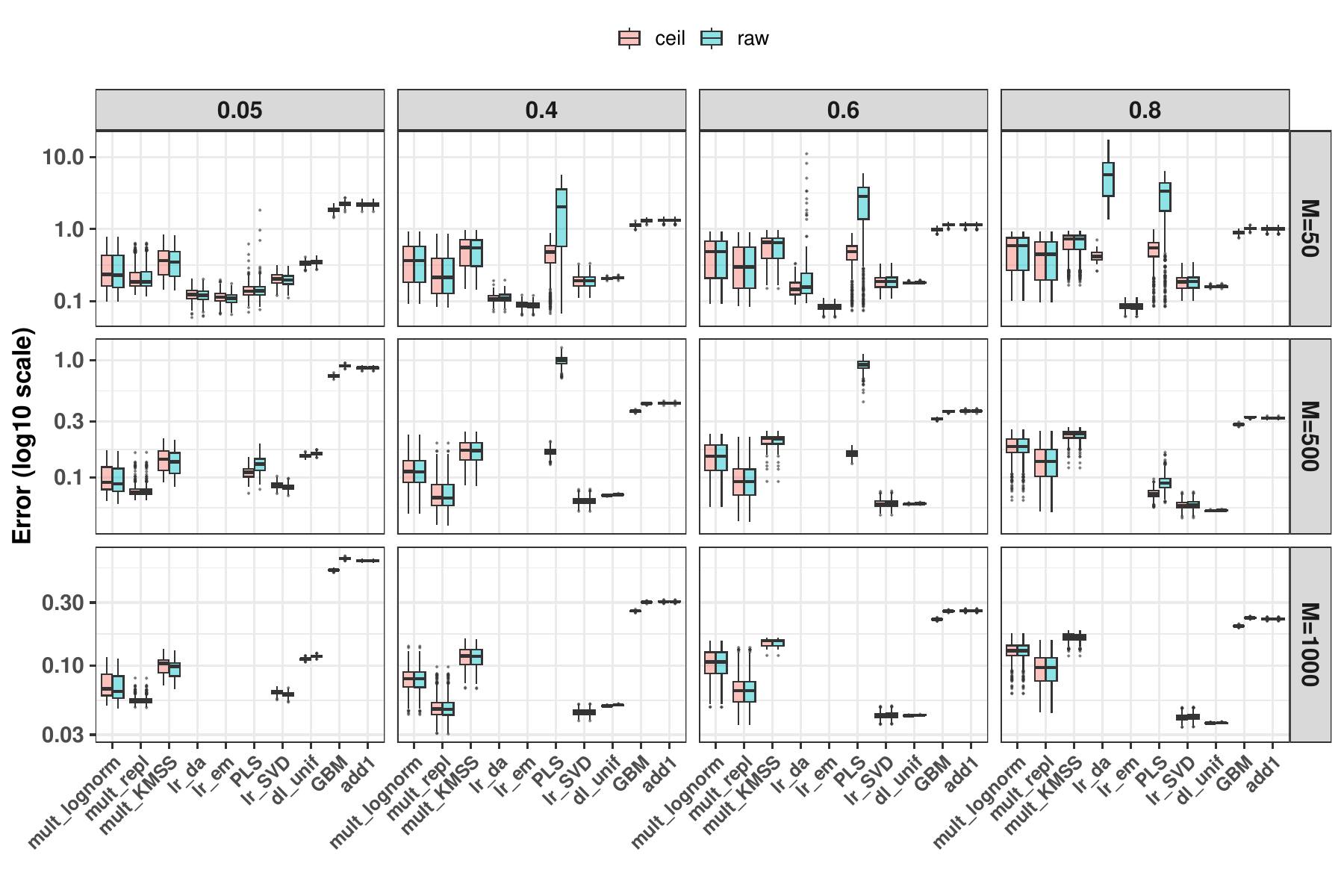}
        \caption{CED}
        \label{fig:CED_boxplot}
    \end{subfigure}

    \vspace{1.2em}

    \begin{subfigure}[b]{0.68\textwidth}
        \centering
        \includegraphics[width=\linewidth]{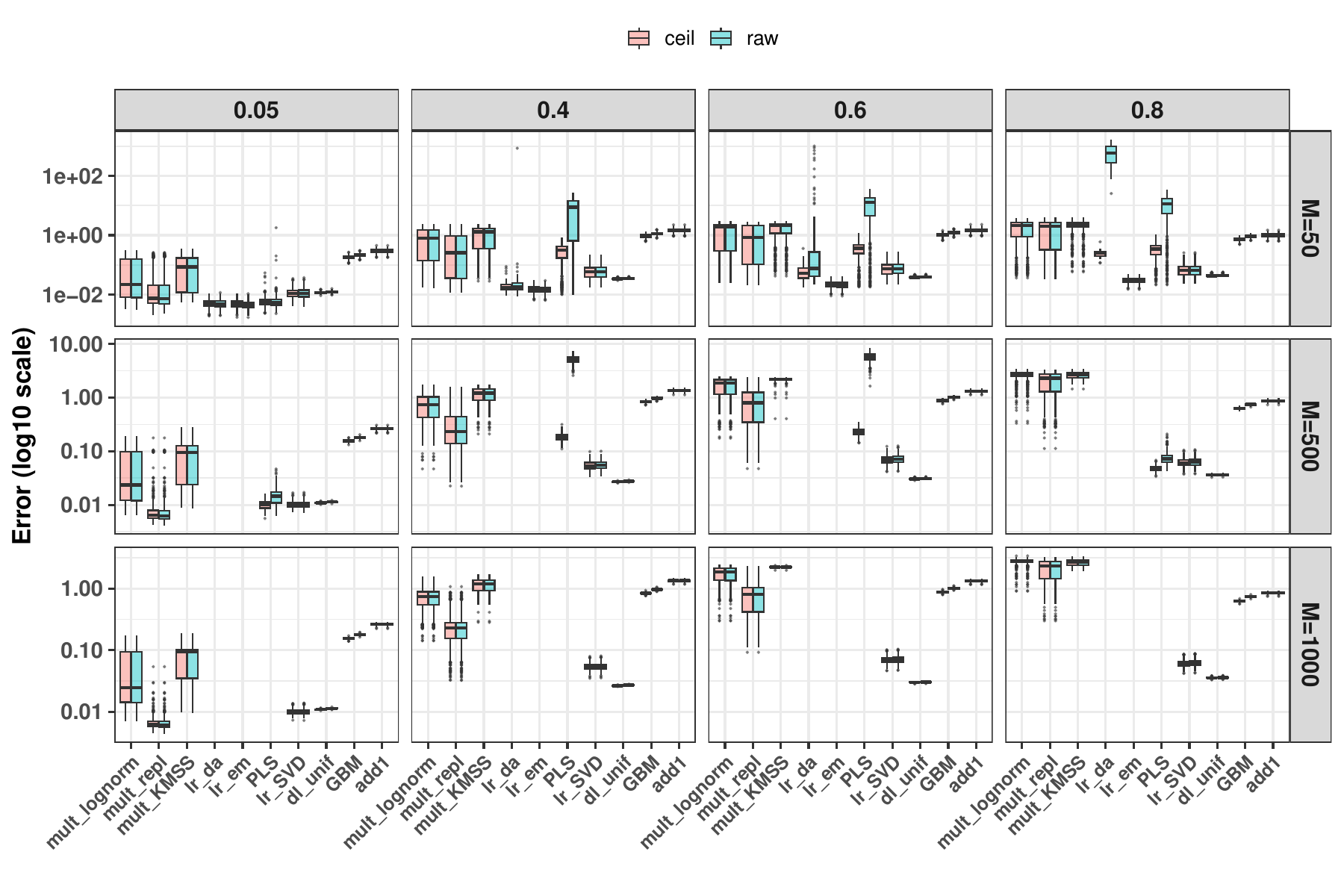}
        \caption{ADCS}
        \label{fig:ADCS_boxplot}
    \end{subfigure}

    \caption{
        Comparison of imputation methods in terms of CED (top) and ADCS (bottom) metrics under
        varying dimensionality ($m = 50, 500, 1000$) and missingness probabilities.
    }
    \label{fig:CED_ADCS_combined}
\end{figure}

\begin{figure}[htbp]
    \centering
    \includegraphics[width=0.95\linewidth]{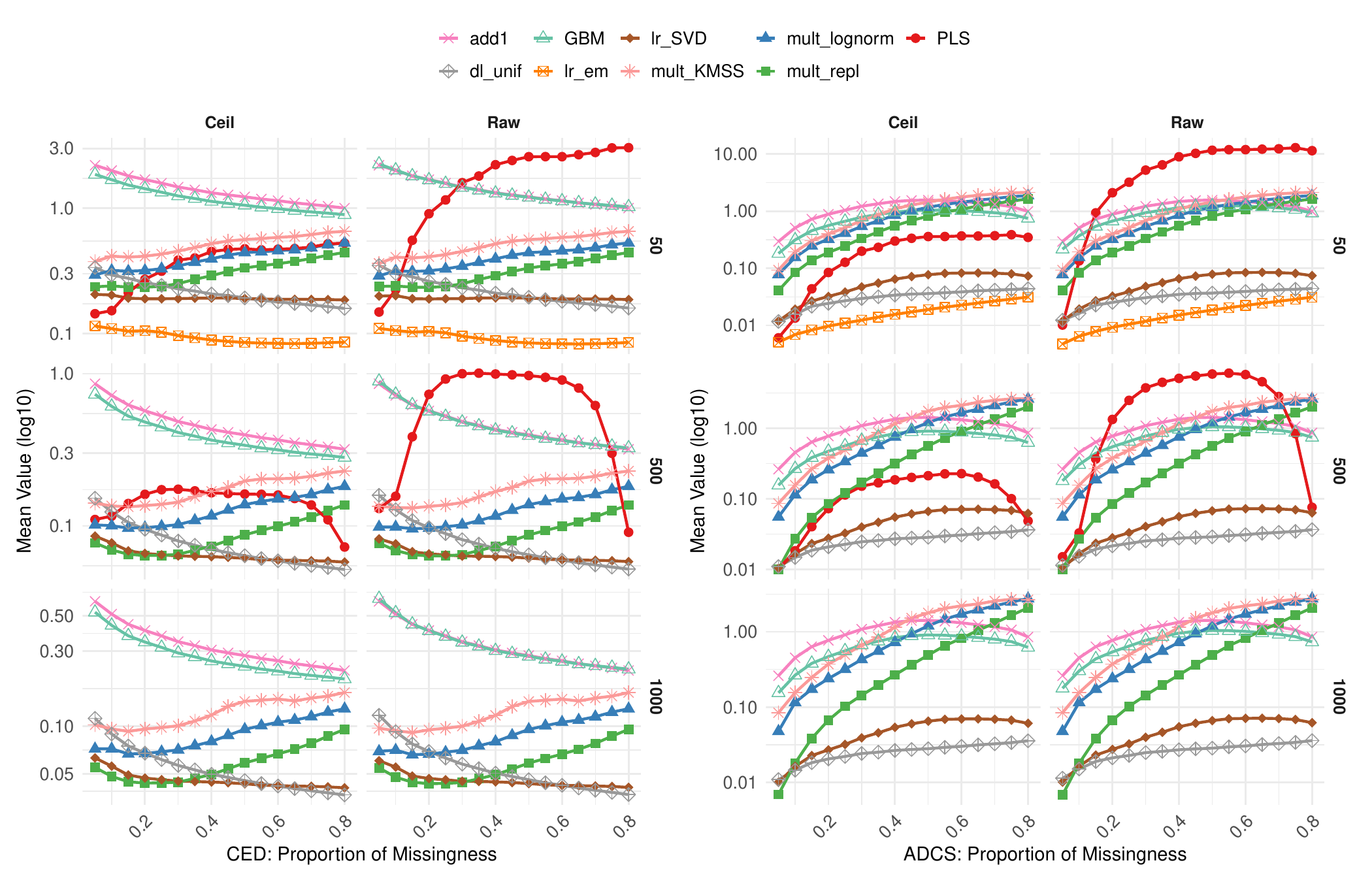}
    \caption{Comparison of imputation methods in terms of CED and ADCS metrics under varying dimensionality and missingness probabilities.}
    \label{fig:CED_ADCS_combined_line}
\end{figure}

\begin{figure}[htbp]
    \centering
    \includegraphics[width=0.95\linewidth]{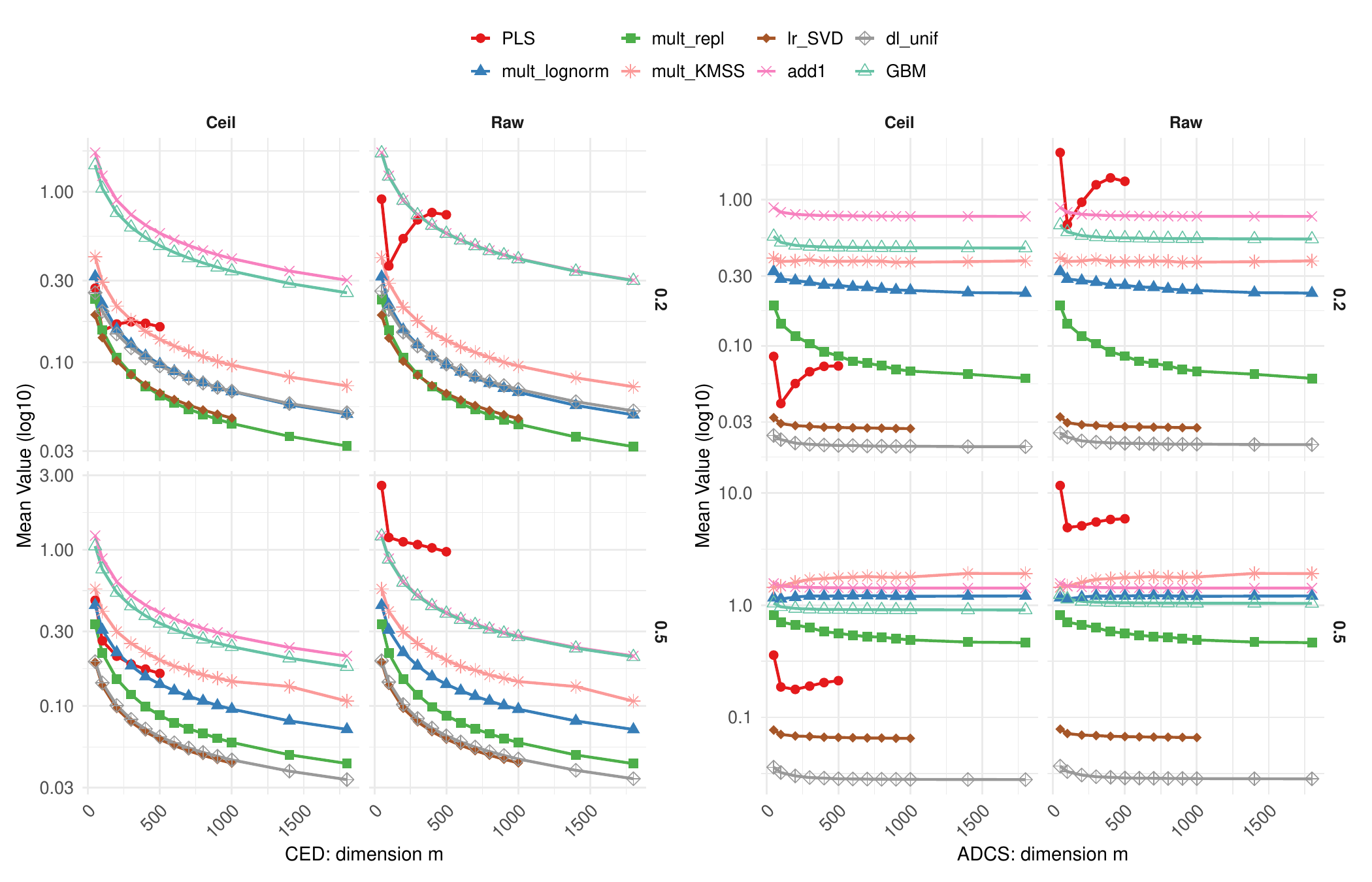}
    \caption{
        Comparison of CED and ADCS metrics (Raw and Ceil)
        across varying dimensions $m$ under fixed missingness probabilities
        $p = 0.2$ and $p = 0.5$.}

    \label{fig:CED_ADCS_combined_p}
\end{figure}

\begin{figure}[htbp]
    \centering
    \includegraphics[width=0.5\linewidth]{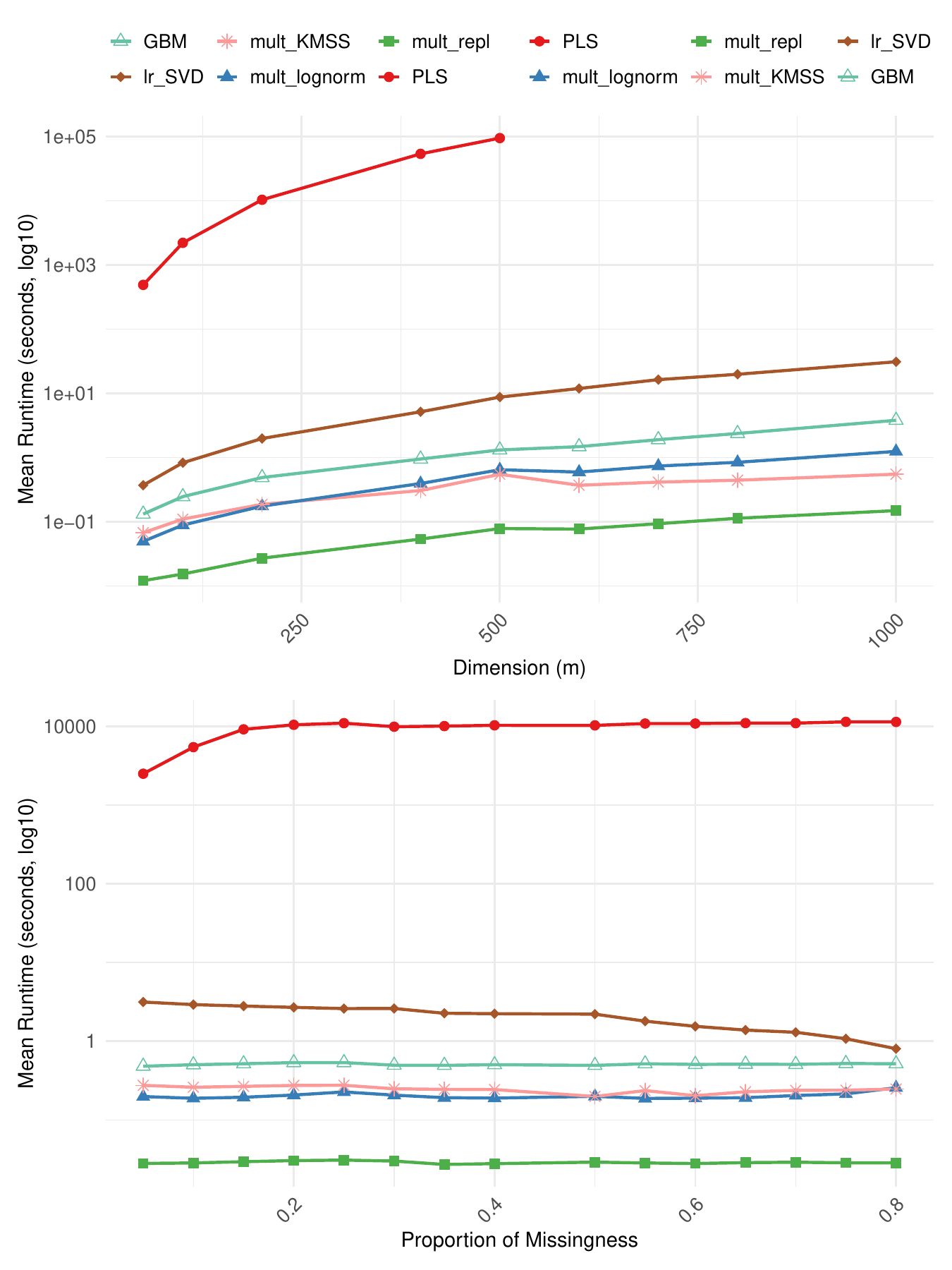}
    \caption{
        Trends of computational runtime under different settings. 
        The upper panel shows the change of mean runtime with increasing dimension ($m$) at fixed missingness probability ($p=0.5$), 
        while the lower panel illustrates the runtime variation with different missingness probabilities ($p$) at a fixed dimension ($m=200$).
    }
    \label{fig:runtime_combined}
\end{figure}

\subsection{Results}

From a technical perspective, the results revealed considerable limitations in several methods.  
For instance, \texttt{lmrob} did not produce valid outcomes in practice.  
The \texttt{lr\_da} method was only applicable in low-dimensional cases (e.g., $m=50$), and even under this setting, the invalid proportion of simulations increased sharply with the proportion of zeros—from 0.28 at $p=0.05$ up to 0.912 at $p=0.8$—indicating low feasibility and a high failure rate.  
Similarly, the \texttt{mult\_KMSS} method also showed increasing failure rates with higher dimensionality: 
while the mean missing rate across all $p$ at fixed $m$ was only 0.056 for $m = 50$, it rose to 0.59 for $m = 1000$.

Figures~\ref{fig:CED_ADCS_combined}–\ref{fig:CED_ADCS_combined_p} summarize performance under combinations of dimensionality and zero proportion. Boxplots were constructed using zero proportions 0.05, 0.4, 0.6, and 0.8 for each dimension, and the average metric values across these levels were used to generate the trend plots in Figs.~\ref{fig:CED_ADCS_combined_line}. These figures illustrate how imputation accuracy changes with increasing missingness across different dimensional settings. In the plots, “raw” denotes the imputed matrices in their original unprocessed (continuous) form, whereas “ceil” indicates that a ceiling operation was applied to the imputed matrices, converting continuous values into integers to better approximate discrete count data.

At the low-dimensional level ($m=50$), the ceiling operation substantially improved the performance of \texttt{PLS} and \texttt{lr\_da} (Fig.~\ref{fig:CED_ADCS_combined}), leading to reduced ADCS and CED values while leaving other methods largely unchanged, with the improvement becoming more pronounced as the zero proportion increased.  
\texttt{GBM}, \texttt{add1}, and \texttt{PLS} exhibited higher CED and ADCS, whereas \texttt{lr\_SVD}, \texttt{lr\_em}, and \texttt{dl\_unif} maintained the lowest deviations.  
The three multiplicative approaches—\texttt{mult\_repl}, \texttt{mult\_lognorm}, and \texttt{mult\_KMSS}—showed moderate, similar growth across zero levels and were mostly unaffected by rounding.

At the moderate-dimensional level ($m=500$), rounding further enhanced \texttt{PLS} performance, particularly at moderate and high missingness levels (Figs.~\ref{fig:CED_ADCS_combined_line}).  
\texttt{mult\_repl} and \texttt{mult\_lognorm} displayed greater variability and occasional instability, while \texttt{GBM} and \texttt{add1} remained less effective.  
\texttt{lr\_SVD} again showed consistently low errors across all levels of missingness.  
Overall, the ceiling operation reduced ADCS fluctuations and improved the stability of methods with initially larger deviations.  
At higher dimensionality (e.g., $m = 1000$), \texttt{lr\_em} and \texttt{lr\_da} no longer produced valid results, reflecting their limited applicability in large-scale settings.

At the high-dimensional level ($m=1000$), \texttt{PLS} failed due to computational constraints, while \texttt{mult\_repl}, \texttt{mult\_lognorm}, and \texttt{mult\_KMSS} exhibited similar increasing patterns in both metrics (Figs.~\ref{fig:CED_ADCS_combined_line}).  
Among these, \texttt{mult\_repl} performed best, achieving the lowest CED at low missingness levels.  
\texttt{lr\_SVD} and \texttt{dl\_unif} remained the most robust, maintaining stable, low errors even under high zero proportions.  
The ceiling operation had little additional benefit at this scale.

Figs.~\ref{fig:CED_ADCS_combined_p} further fix the missingness proportions at $p=0.2$ and $p=0.5$ to explore how dimensionality affects performance.  
The trends are largely consistent with previous findings: the upward rounding (ceiling) substantially improves \texttt{PLS} performance, especially at moderate-to-high zero levels; the three multiplicative imputation methods perform well in CED when the proportion of zeros is low—indicating better preservation of the original compositional geometry—but perform poorly in ADCS due to induced artificial correlations.  
\texttt{lr\_SVD} and \texttt{dl\_unif} again demonstrate stable behavior across dimensions.  
Figure~\ref{fig:runtime_combined} presents the computational runtime comparison, showing that \texttt{PLS} is the most time-consuming method among all. The computations were performed using the Puhti supercomputer provided by CSC – IT Center for Science, Finland.

To further explain the substantial improvement observed for the \texttt{PLS} method after applying the ceiling operation, we conducted a localized, fine-grained analysis under different combinations of dimensionality and zero proportions. Although each simulation setting was repeated 500 times before, the purpose of this part is to examine the detailed imputation behavior, which can already be clearly illustrated by inspecting a single replicate. Therefore, for illustrative and explanatory purposes, we present one representative replicate (\texttt{rep = 1}) under a moderate-dimensional setting ($m=200$) and a zero proportion of $p=0.2$. This focused examination enables us to visualize the underlying mechanics of the methods without loss of generality. For interpretability, six representative columns were selected—two each from the low-, medium-, and high-abundance groups. Boxplots of imputed values were produced for all methods, and scatter plots comparing true ($x$) and imputed ($y$) values were generated for \texttt{PLS} and \texttt{lr\_SVD} (Fig.~\ref{fig:box_scatter_combined_p020_m200}).

As shown in the scatter plots, although the overall evaluation metrics indicate weaker performance for \texttt{PLS (raw)}, see Fig. \ref{fig:CED_ADCS_combined}, most imputed values lie close to the reference line $y=x$, indicating that the majority of estimates are near their true counterparts. One entire column of imputed values exhibited extremely small magnitudes, with \texttt{mg820} clearly identifiable as the variable responsible for this outlier pattern in this replicate. After applying the ceiling operation, these very small continuous values were converted into integers, leading to a substantial reduction in both CED and ADCS.This example illustrates the mechanism behind the metric degradation using a single replicate and six representative columns; similar outlier behavior may also occur in the remaining unshown columns or other repetitions. 

In addition, due to the wide dynamic range of the rabbit dataset—spanning several orders of magnitude—the multiplicative methods (\texttt{mult\_repl}, \texttt{mult\_lognorm}, and \texttt{mult\_KMSS}) occasionally produced rows (samples) containing all-negative imputed values.

\subsection{Discussion and recommendations}

In summary, since many imputation methods were originally developed for continuous compositional data, applying them directly to discrete count data requires caution due to their sensitivity to sparsity, scale differences, and rounding. Future work should focus on developing methods that explicitly account for the discrete-based nature of the data, providing a more practical and reliable bridge between log-ratio theory and real high-dimensional sequencing count data.

The observed patterns highlight several important methodological considerations for applying imputation algorithms to compositional count data. Log-ratio–based methods such as \texttt{lr\_SVD} and \texttt{lr\_em} consistently demonstrated strong robustness across dimensionalities and zero proportions. 
The multiplicative replacement approaches (\texttt{mult\_repl}, \texttt{mult\_lognorm}, and \texttt{mult\_KMSS}), while computationally fast and still performing reasonably when the zero proportion is low, showed clear instability under high sparsity or heterogeneous count scales. In particular, the occasional occurrence of all-negative rows generated by these methods can be attributed to numerical instabilities that arise when log-ratio transformations interact with the large dynamic range and left-censored components of the Rabbit dataset—a limitation noted earlier when these methods were introduced.

\begin{figure}[htbp]
    \centering
    \includegraphics[width=\textwidth]{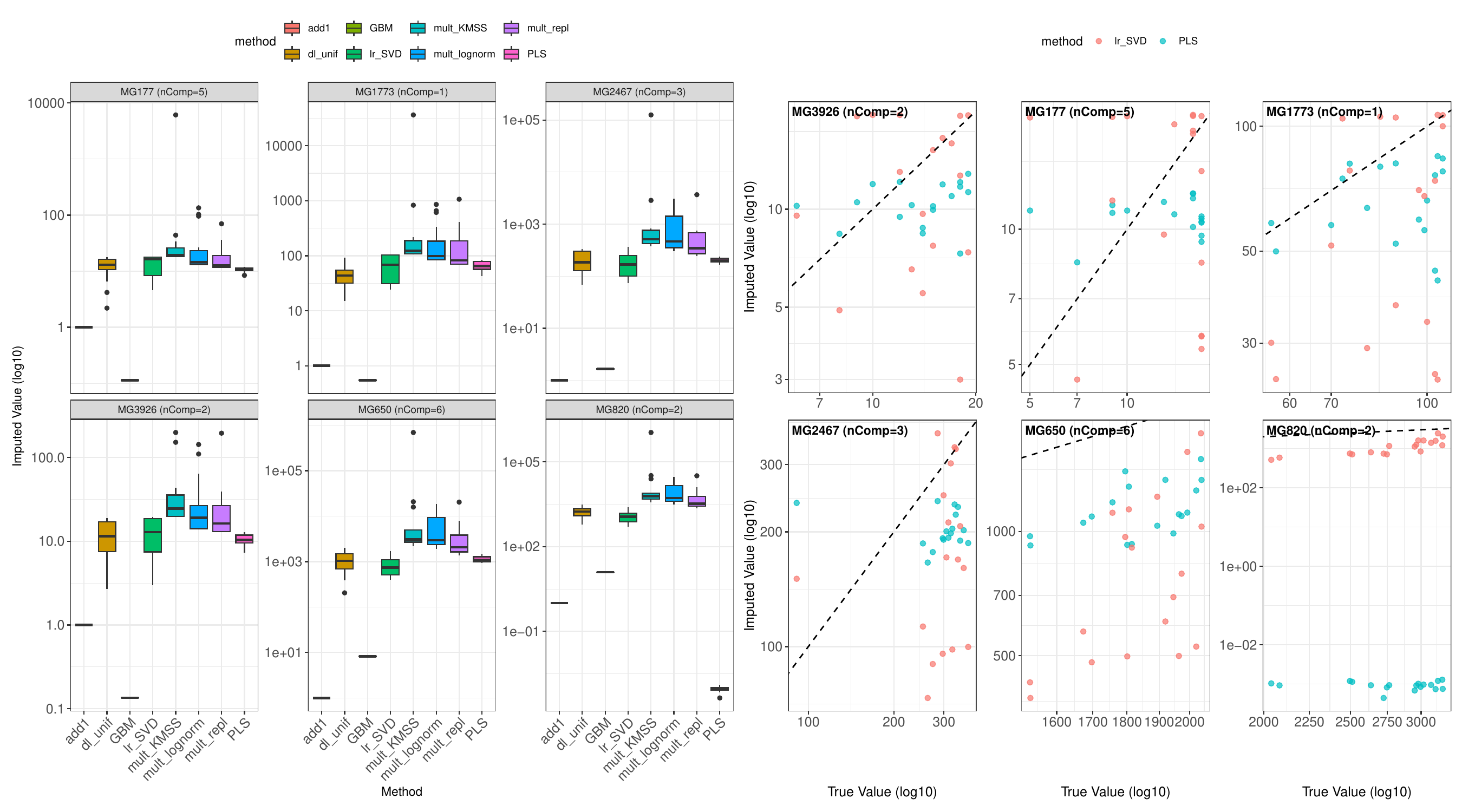}

    \caption{
              Local inspection of imputation behavior under $m=200$, $p=0.2$, and $\mathrm{rep}=1$. 
        The left portion of the figure shows the distribution of imputed values across 
        different imputation methods for six representative variables (two from each 
        abundance level: low, medium, and high), with the corresponding number of PLS 
        components ($n_{\mathrm{comp}}$) annotated in each facet.  
        The right portion displays scatter plots comparing true versus imputed values 
        (with $y = x$ as a reference line) for the \texttt{PLS} and \texttt{lr\_SVD} 
        methods.  While most \texttt{PLS} imputations closely follow the identity line, 
        certain variables (e.g., \texttt{mg820}) show pronounced deviations, which in turn 
        lead to inflated log-ratio–based error measures such as CED and ADCS in global 
        evaluations.
    }
    \label{fig:box_scatter_combined_p020_m200}
\end{figure}

The localized example further illustrates the mechanism underlying the weaker aggregated performance of \texttt{PLS} in its raw form. Although the global evaluation metrics (CED and ADCS) suggest reduced accuracy, most imputed values from \texttt{PLS (raw)} align closely with the reference line $y = x$, indicating strong preservation of proportional relationships. The degradation was instead driven by rare but extremely small imputed values—such as those observed for variable \texttt{mg820} in Figure\ref{fig:box_scatter_combined_p020_m200}—which produced entire columns of near-zero estimates in specific replicates. Because CED and ADCS operate within the log-ratio form, these tiny positive values are disproportionately amplified, inflating the error scores. Applying a ceiling operation corrects these values by converting them to integers, effectively mitigating logarithmic amplification, consistent with the earlier discussion that quantization error has the strongest impact on extremely small values—particularly those below 1—where values far less than 1 are rounded up to 1 and thus the extreme distortions in the log-ratio computations are greatly alleviated. This reduction in error therefore reflects decreased metric sensitivity rather than an inherent improvement in the underlying predictive accuracy. Similar rare-value effects may occur in the remaining variables or across different repetitions, even though they are not all shown here. A supplementary comparison study based on simulated count data is provided in Appendix~\ref{appendix.A}, further confirming that the observed behavior of \texttt{PLS} is not specific to the Rabbit dataset but reflects a more general mechanism. Additional examples illustrating how such rare small-value artifacts arise under other dimensionalities and zero proportions are included in Appendix~\ref{appendix.B}.

Building on these observations, a broader methodological implication emerges. Our results indicate that directly applying continuous CoDA frameworks to discrete count data can introduce systematic bias arising from quantization error and scale disparity, particularly under low-count or highly heterogeneous conditions. Moreover, because the log-ratio framework inherently assumes strictly positive and continuous inputs, its use on discrete counts further amplifies distortions through discretization noise and altered correlation structures. These issues become especially pronounced when extremely small values are present, as discussed earlier. Consequently, when employing log-ratio–based evaluation metrics such as CED or ADCS on count-valued compositional data, particular caution is warranted, and the potential effects of data discreteness should be explicitly accounted for during interpretation. In addition, our experiments show that applying a ceiling operation to the raw imputed values—thereby restoring the discrete nature of the data—can mitigate some of these distortions and improve the effectiveness and stability of certain methods.

From a practical perspective, method selection must balance accuracy, robustness, and computational cost. Multiplicative replacement methods may be adequate when zero proportions are small and scale variation is moderate, but their instability in heterogeneous environments limits their reliability. \texttt{lr\_SVD} consistently delivered stable performance across all tested dimensions and sparsity levels, making it a dependable default choice. Although \texttt{PLS} is computationally expensive, it retains high predictive fidelity in most settings and remains valuable when computational resources permit. In practice, however, its raw outputs should be inspected for rare small-value artifacts, and rounding should be applied when necessary to ensure numerical stability. The \texttt{dl\_unif} method often achieves favorable aggregated metric scores but offers limited interpretability due to its stochastic nature, suggesting its use primarily as a baseline or supplementary approach. More generally, modestly preserving the discrete nature of the data—for example, via rounding or integer mapping—can improve both stability and interpretability in downstream analyses.

In summary, while continuous CoDA methods form the basis of most existing imputation algorithms, their direct application to sparse or high-dimensional count data requires careful consideration of scale heterogeneity, zero inflation, and rounding effects. Future work should focus on developing imputation frameworks that explicitly incorporate the discrete and lattice-valued nature of count data, thereby providing a more coherent bridge between log-ratio methodology and the unique structure of sequencing-based compositional data. Such approaches hold promise for improving both theoretical soundness and practical performance in high-dimensional compositional count settings.

\section{Conclusion}
In this review, we examined the challenge of zero values in compositional data, motivated in particular by high-dimensional and extremely sparse sequencing-based microbiome datasets. As sequencing technologies continue to advance, such data have become increasingly common, yet conventional log-ratio–based compositional data analysis remains fundamentally incompatible with zeros. The problem becomes even more complex in sequencing count data, where measurements are discrete and lattice-valued rather than continuous, violating key assumptions underlying classical CoDA.
We began by revisiting the geometric foundations of compositional data and emphasized the importance of distinguishing between rounded zeros and essential zeros, as they arise from different mechanisms and demand different treatments. Building on this distinction, we summarized three major methodological directions: zero-tolerant transformations, zero-replacement strategies that construct strictly positive compositions, and model-based approaches that retain zeros as statistically meaningful quantities. We further reviewed extensions specifically developed for count-based compositional data, elaborating on why their discrete nature is fundamentally incompatible with classical continuous CoDA, and summarising current strategies tailored to sparse count compositions, including probabilistic modelling frameworks such as Dirichlet–multinomial and zero-inflated formulations, as well as imputation methods designed for discrete data.

Nonetheless, while we recognise the theoretical value of probabilistic modelling frameworks—such as Dirichlet–multinomial formulations—and the usefulness of zero-tolerant transformations, the primary aim of this study was different. Our goal was to systematically evaluate the practical performance and general applicability of existing mainstream matrix-based imputation strategies—most of which were originally designed under the log-ratio framework—when applied to high-dimensional, sparse HTS count data, without presupposing any specific statistical model or probabilistic structure. By focusing on methods that operate directly on the data matrix, we aim to provide HTS researchers with a pragmatic and model-agnostic guide for tool selection, emphasising numerical accuracy and structural fidelity rather than the fit of any particular statistical model.

Based on these premises, we sought to assess how well current imputation methods can accommodate the distinctive properties of sequencing-derived count compositions, and to offer practical guidance for researchers analysing such data. To this end, we conducted comparative experiments on both real high-dimensional datasets and simulated data generated from real microbial compositions. The results show that while some methods—particularly log-ratio–based approaches such as \texttt{lrSVD} and \texttt{lrEM}—exhibit strong robustness, most existing imputation strategies are designed for continuous data and fail to respect the discrete nature of sequencing counts. Recovering a count-like structure typically requires an additional ceiling step, and our results show that this post-processing can substantially improve performance for certain methods, especially \texttt{PLS}, whose raw outputs may contain extremely small values that become highly distorted in log-ratio space. In our comparison, we therefore applied ceiling adjustments to obtain discrete outputs, suggesting that future research could further explore which post-processing strategies are most effective in stabilizing imputations for sparse count compositions.
Nonetheless, our matrix-level evaluation—based on both global covariance-based metrics and feature-wise comparisons—demonstrates that it remains difficult to achieve imputations that are simultaneously close to the true values and structurally faithful to the original count matrix. Methods that perform well on average may still break down on individual components, and no existing approach consistently dominates across numerical accuracy and overall evaluation criteria.

This reveals a central gap in current research: existing zero-handling strategies either rely on continuity assumptions or inadvertently break the lattice structure of count data, often requiring post-processing. There is a clear need for new methodological development that jointly addresses zero inflation, compositional constraints, and discreteness within a coherent statistical framework.
Looking ahead, promising directions include unified frameworks that bridge imputation with count-based probabilistic models, ensuring that methods respect the discrete nature of sequencing data. At the same time, log-ratio analysis itself may require refinement when applied to count data—for example, by incorporating safeguards for extremely small values or adapting interpretation strategies depending on count scale. Finally, progress will also depend on the development of standardized benchmarks, reproducible workflows, and accessible software implementations.
By organizing the scattered literature and clarifying the statistical nature of zeros, this review establishes a more coherent conceptual framework for zero handling in compositional settings. This perspective helps connect the well-established log-ratio methodology with the growing need to analyze discrete, sequencing-based count data, where zeros are frequent. We believe this work provides a solid basis for future advances, including methods explicitly designed for high-dimensional compositional count data.
\section*{Acknowledgments}
This work was supported by the European Commission grant no. CZ.02.01.01/00/23\_025/008686,and by the Finnish Doctoral Program Network in Artificial Intelligence (AI-DOC; decision no. VN/3137/2024-OKM-6). We also acknowledge CSC – IT Center for Science, Finland, for providing computational resources.

\section*{Declaration of generative AI and AI-assisted technologies in the manuscript preparation process}
During the preparation of this work, the author(s) used AI tools to improving the clarity and readability of the manuscript. After using this tool, the author(s) reviewed, edited, and verified the content as needed, and take full responsibility for the content of the published article.

\appendix
\section{Simulation design for zero-imputation comparison }
\label{appendix.A}

In this appendix, we present the results of applying the imputation methods listed in Table\ref{sim:imputation table} to a simulation dataset.

\subsection{Construction of zero-free simulation dataset}

We begin with a real microbiome count matrix \(\mathbf{Y}\), taken from the \texttt{gllvm} example dataset. The data consist of normalized read counts of 985 bacterial species sampled from 56 soil sites across three distinct climatic regions: Ny-\AA lesund (high Arctic), Kilpisj\"arvi (low Arctic), and Mayrhofen (European Alps) \cite{Kumar2017}. Species were identified based on 16S rRNA similarity and clustered into OTUs. The dataset exhibits 63.3\% sparsity (zeros).

Following the procedure in \cite{Fernandes2013}, we construct a corresponding zero-free dataset using a Bayesian smoothing step.

For each sample \(i\) with observed count vector \(\mathbf{n}_i\), we place a symmetric Dirichlet prior with parameter \(\tfrac{1}{2}\) and draw from the posterior:
\[
\mathbf{p}_i \mid \mathbf{n}_i \sim \text{Dirichlet}\!\left(\mathbf{n}_i + \tfrac{1}{2}\right).
\]
Collecting all samples yields a strictly positive compositional matrix:
\[
P_{\text{post}} = \{ \mathbf{p}_i \}_{i=1}^n.
\]
This assigns strictly positive mass to all taxa, including those originally observed as zero.

To obtain a count matrix analogous to sequencing output, we scale the posterior compositions by a fixed sequencing depth that was manually increased to ensure that rounding would not introduce zeros:
\[
Y_{\text{nozero},ij} = \text{Round}\!\bigl( P_{\text{post},ij} \cdot \text{depth}_{\text{full}} \bigr).
\]
This yields a fully positive integer count matrix of dimension \(56 \times 985\) 
that contains no zeros, preserves key characteristics typical of high--throughput 
sequencing data---such as a wide dynamic range---similar to those observed in the 
Rabbit dataset, and retains its inherent compositional nature, making it suitable 
for log--ratio--based evaluation.

This Dirichlet-based construction followed by scaling is consistent with simulation strategies used in microbiome studies, including \cite{ChenLi2013DMR}.

The zero-insertion and imputation experiments follow the design in Section \ref{sec:Experiment_design}, using \(m = 50\) and \(m = 200\) with \(p = 0.2\), due to computational constraints. We note this briefly here, as the experimental logic otherwise remains unchanged.
\subsection{Result}
\begin{figure}[htbp]
    \centering
    \includegraphics[width=0.95\linewidth]{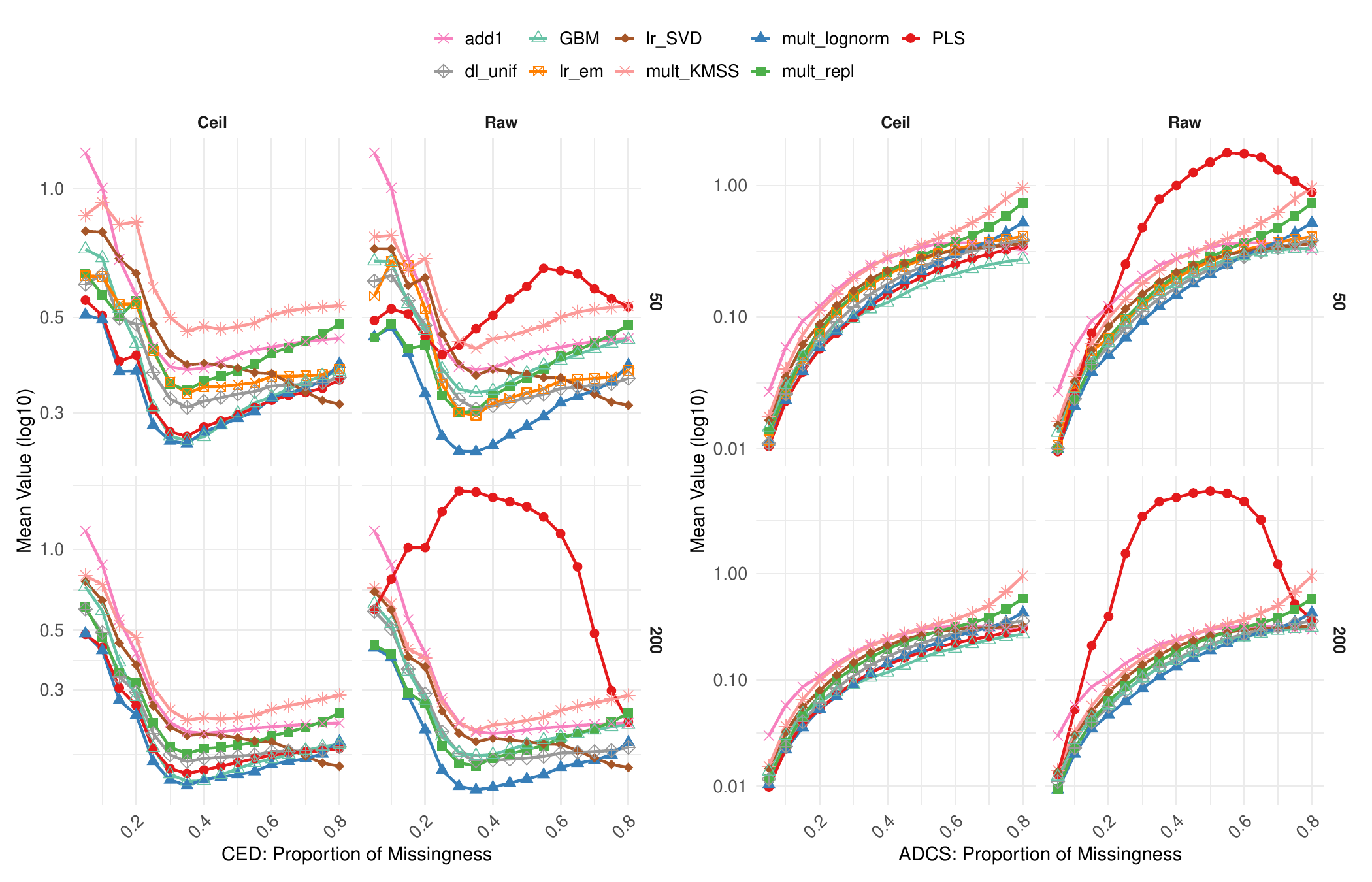}
    \caption{Comparison of imputation methods in terms of CED and ADCS metrics under varying dimensionality and missingness probabilities with simulation dataset.}
    \label{fig:CED_ADCS_combined_sim}
\end{figure}

\begin{figure}[htbp]
    \centering
    \includegraphics[width=0.95\linewidth]{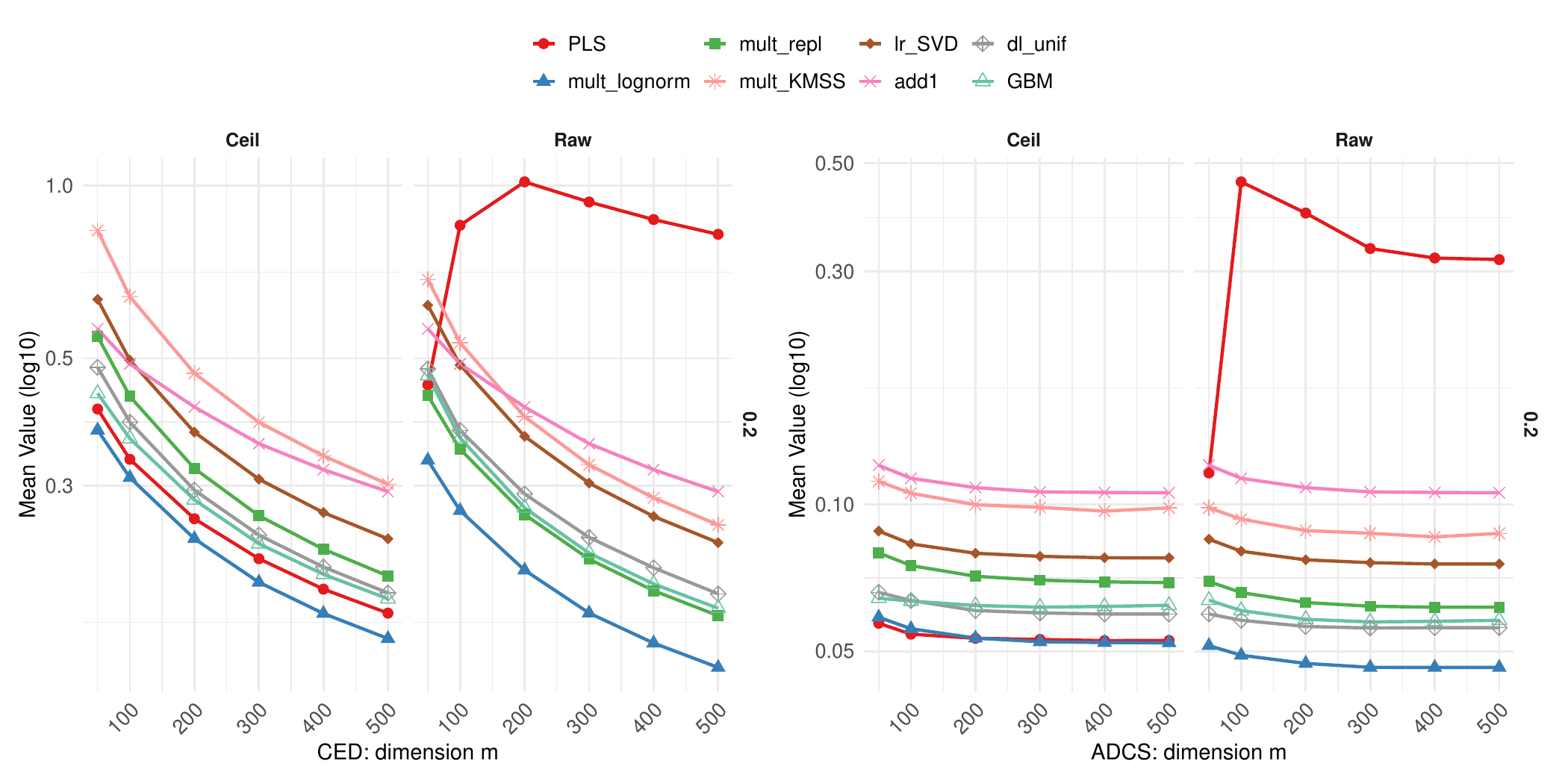}
    \caption{
        Comparison of CED and ADCS metrics (\textit{Raw} and \textit{Ceil})
        across different dimensions $m$ under a fixed missingness probability
        of $p = 0.2$ under simulation dataset. 
    }
    \label{fig:CED_ADCS_combined_p02_sim}
\end{figure}

\begin{figure}[htbp]
    \centering

    \begin{subfigure}[b]{0.68\textwidth}
        \centering
        \includegraphics[width=\linewidth]{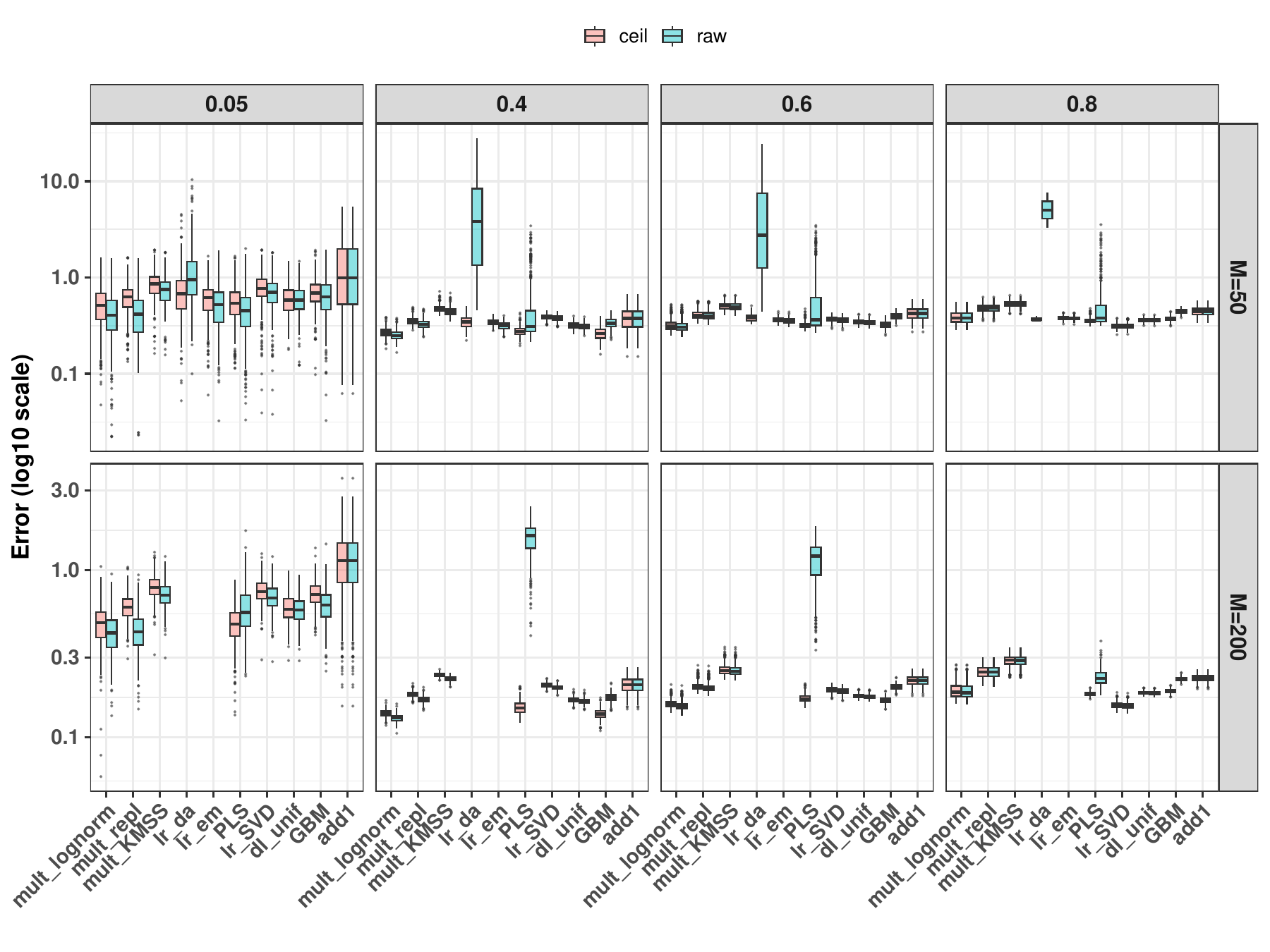}
        \caption{CED}
        \label{fig:CED_boxplot_sim}
    \end{subfigure}

    \vspace{1.2em}

    \begin{subfigure}[b]{0.68\textwidth}
        \centering
        \includegraphics[width=\linewidth]{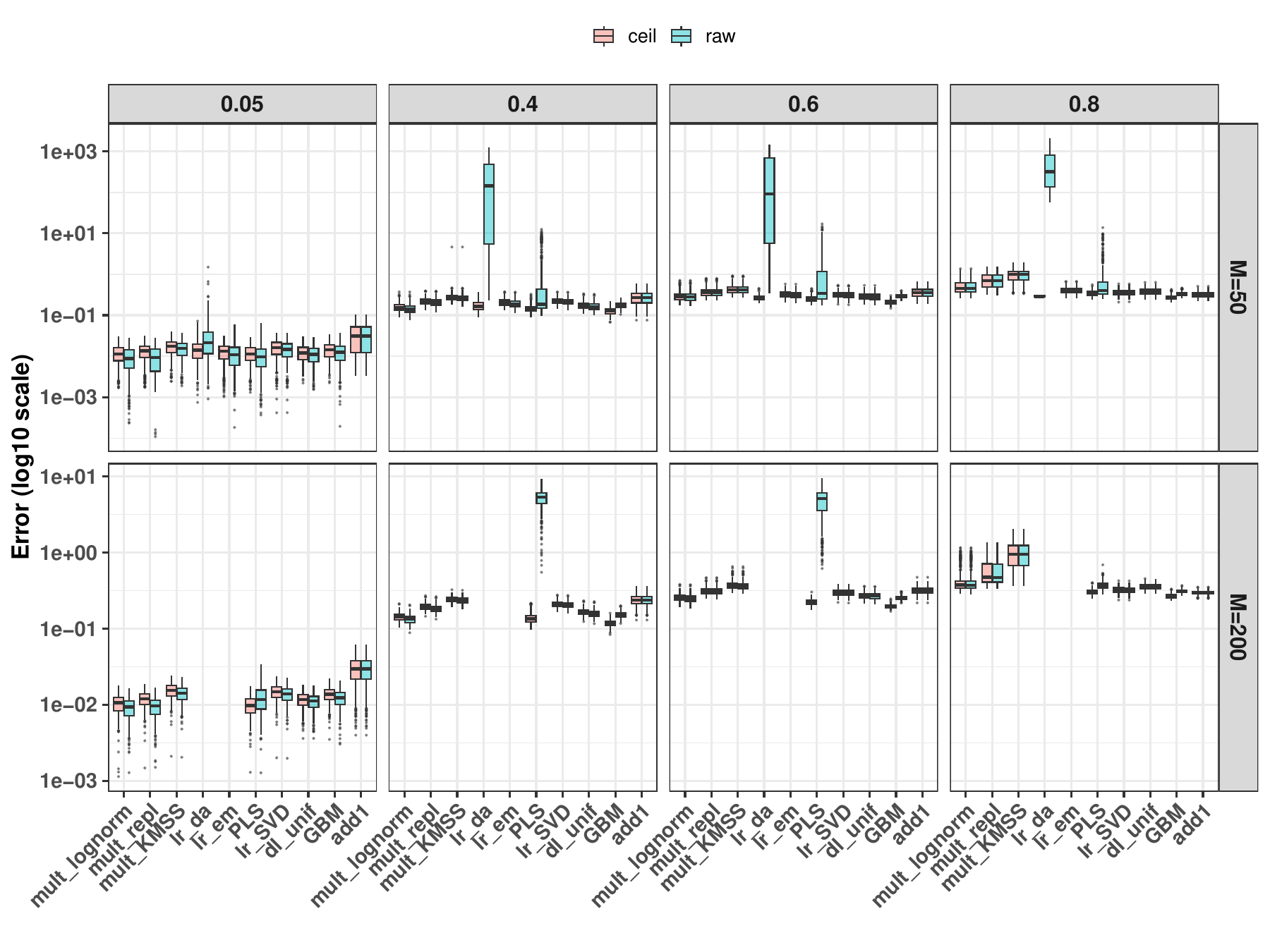}
        \caption{ADCS}
        \label{fig:ADCS_boxplot_sim}
    \end{subfigure}

    \caption{
        Simulation-based comparison of imputation methods in terms of CED (top) 
        and ADCS (bottom) metrics across dimensionality ($m = 50, 200$) and missingness probabilities.
    }
    \label{fig:CED_ADCS_combined_sim_box}
\end{figure}

Across all simulation scenarios, the results consistently show that the behavior of
\texttt{PLS} after applying the ceiling operation closely parallels the patterns
observed in the main experiment. As shown in
Figures~\ref{fig:CED_ADCS_combined_sim}--\ref{fig:CED_ADCS_combined_sim_box}, the
ceiling adjustment substantially reduces the CED and ADCS values, resulting in lower
and more stable metric outputs. Because the Rabbit dataset is representative of
HTS-derived count data—with discrete support, heterogeneous scales, and a wide dynamic
range—the simulation data were constructed to share these structural characteristics.
The consistent improvement of \texttt{PLS} after rounding across both real and
simulated datasets indicates that this behavior reflects a general pattern in
count-valued compositional settings, rather than an effect specific to any particular
dataset.

\section{Additional diagnostic figures}
\label{appendix.B}
To complement the localized example discussed in the main text, this appendix provides 
additional local diagnostic plots that follow the same experimental design as in the 
main body, with the same data-generating mechanism and evaluation procedure, but using 
different combinations of dimensionality ($m$) and zero proportion ($p$). These plots 
illustrate how rare but extremely small imputed values can arise under various settings, 
even when only $m$ and $p$ are altered.

These examples serve two purposes. First, they demonstrate that the appearance of 
near-zero imputation artifacts generated by \texttt{PLS} is not limited to the specific 
case highlighted in the main manuscript (e.g., variable \texttt{mg820}), but also occurs 
across a broader range of simulation settings. Second, they show that the presence of 
such extremely small values can affect the stability of log-ratio--based evaluation 
metrics, suggesting that this issue deserves further attention when log-ratio methods 
are applied to count data. This phenomenon may arise more broadly in count-based 
compositional settings and does not depend on the particular variable or dimensional 
configuration involved.

\begin{figure}[htbp]
    \centering

    \begin{subfigure}[t]{0.95\textwidth}
        \centering
        \includegraphics[width=\linewidth]{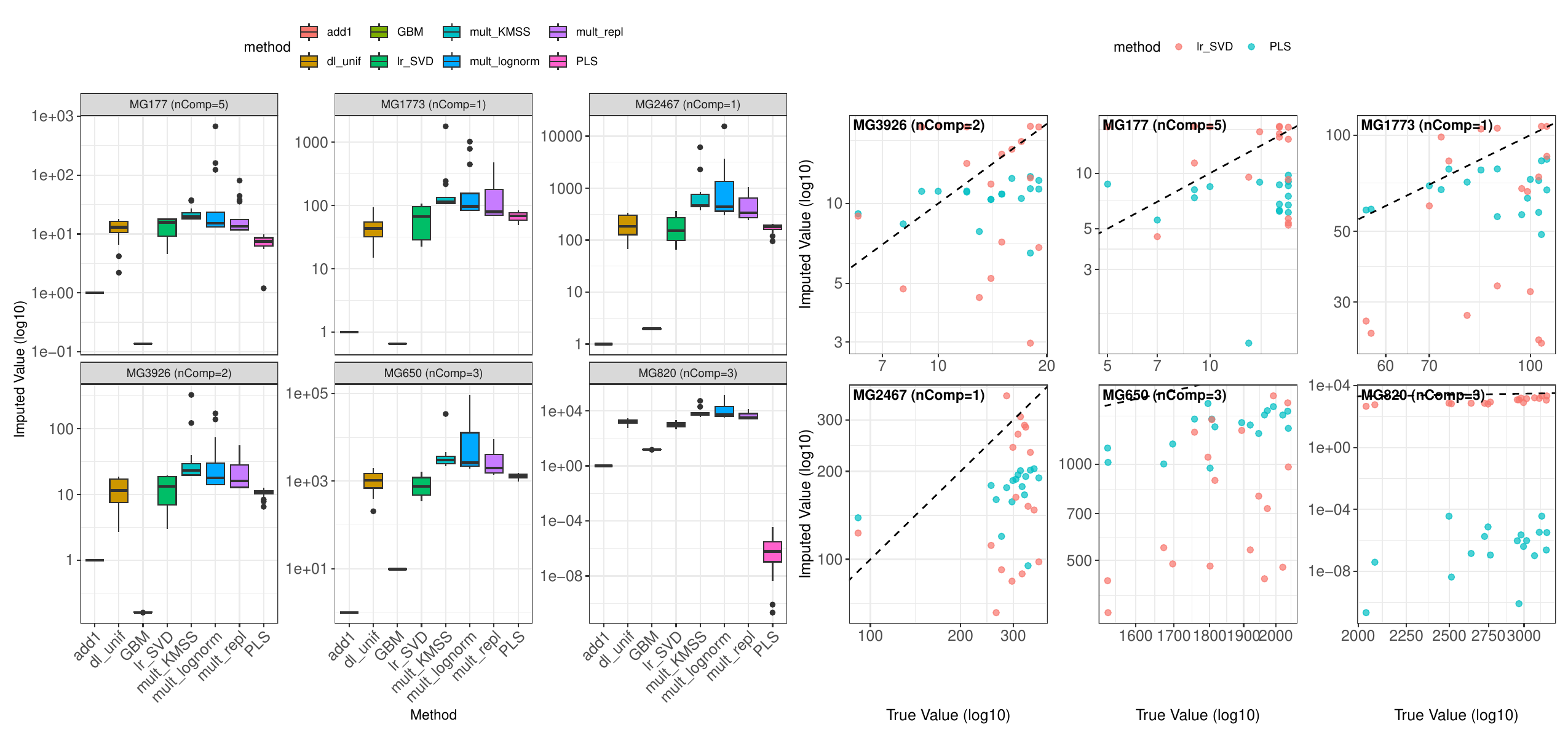}
        \caption{
             $m = 100$, $p = 0.20$,  
            Abnormal variable \texttt{MG820}.
        }
        \label{fig:appendix_p020_m100}
    \end{subfigure}

    \vspace{1.6em}

    \begin{subfigure}[t]{0.95\textwidth}
        \centering
        \includegraphics[width=\linewidth]{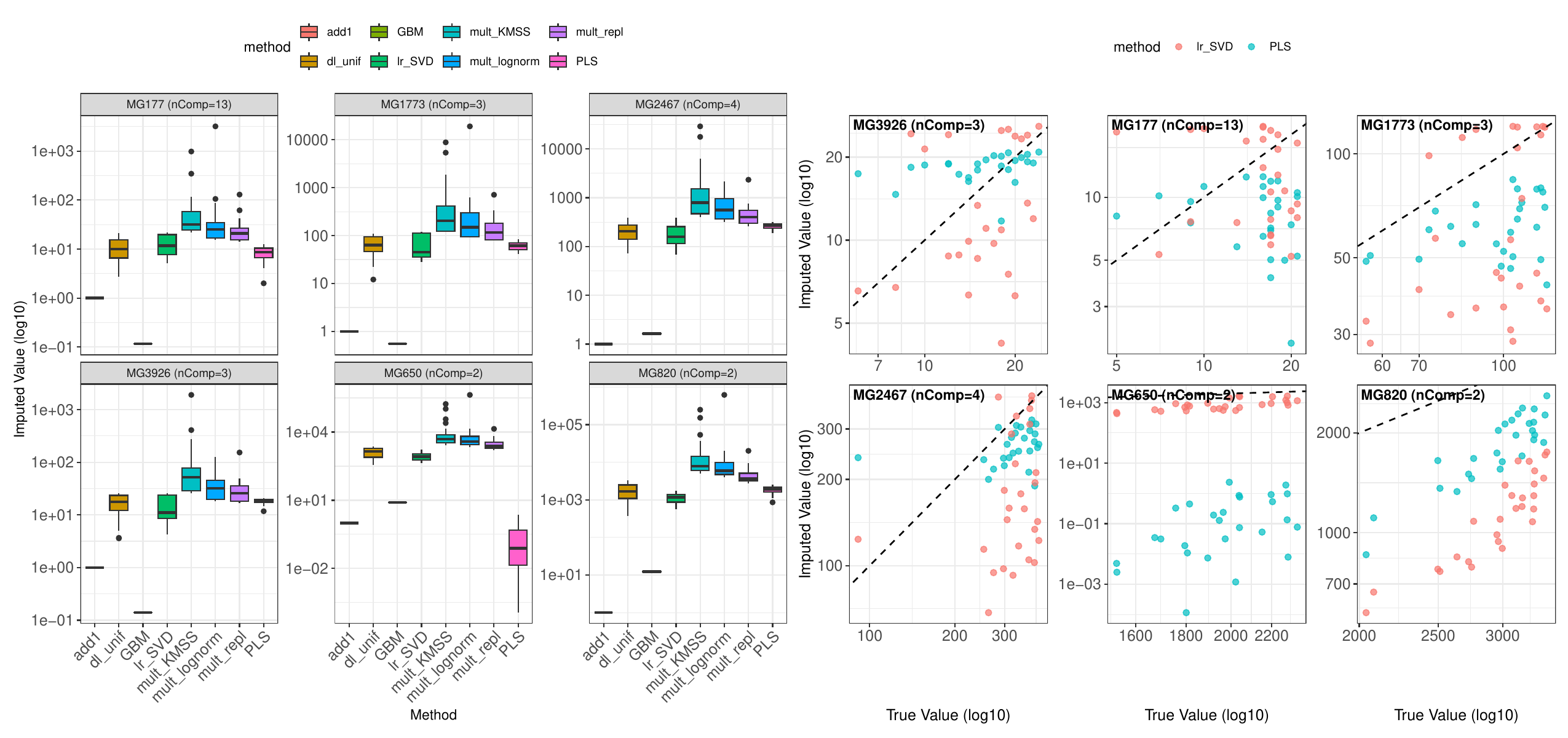}
        \caption{
             $m = 500$, $p = 0.30$,  
            Abnormal variable \texttt{MG650}.
        }
        \label{fig:appendix_p030_m500}
    \end{subfigure}

    \caption{(a) corresponds to $m = 100$, $p = 0.20$ with abnormal variable \texttt{MG820},  
and (b) corresponds to $m = 500$, $p = 0.30$ with abnormal variable \texttt{MG650}.  
Each panel summarizes the distribution of imputed values across methods together with 
true–imputed comparisons for representative     variables.}

    \label{fig:appendix_combined_three}
\end{figure}

Figures~\ref{fig:appendix_p020_m100} and \ref{fig:appendix_p030_m500} present two 
representative instances showing that, under different dimensionalities and 
zero-proportion settings, a single variable can again produce extremely small raw 
imputations under \texttt{PLS}, resulting in entire columns of near-zero values.

\bibliographystyle{apalike}  
\bibliography{references}

\end{document}